\documentclass[11pt,oneside,letterpaper]{article}
\pdfoutput=1

\usepackage{amssymb}
\usepackage{amsmath}
\usepackage[dvips]{graphicx}
\usepackage{setspace}
\usepackage{fancyhdr}
\usepackage{xcolor}
\usepackage{cite}
\usepackage{ifpdf}
\usepackage{graphicx}
\usepackage{rotating}
\usepackage{comment}
\usepackage{geometry}
\usepackage[USenglish]{babel}

\usepackage{color}
\definecolor{darkgreen}{rgb}{0,0.5,0}
\definecolor{darkblue}{rgb}{0,0,0.6}
\definecolor{purple}{rgb}{0.4,.2,0.7}
\definecolor{awesome}{rgb}{1.0, 0.13, 0.32}

\usepackage[colorlinks=true,citecolor=darkgreen,linkcolor=purple,urlcolor=purple]{hyperref}

\newcommand{\bi}{\begin{itemize}}
\newcommand{\ei}{\end{itemize}}
\newcommand{\bea}{\begin{eqnarray}}
\newcommand{\eea}{\end{eqnarray}}
\newcommand{\be}{\begin{equation}}
\newcommand{\ee}{\end{equation}}

 \definecolor{olivegreen}{rgb}{0,0.52,0.17}
 \def\TA#1{{\color{olivegreen}{[#1 -TA]}}}

\numberwithin{equation}{section}
\allowdisplaybreaks  

\begin{document}

\vspace*{2.5cm}
\begin{center}
{ \LARGE  \textsc{Marginal Deformations $\&$ Rotating Horizons}}
\\
\vspace*{1.7cm}
Dionysios Anninos$^1$, Tarek Anous$^2$ and Raffaele Tito D'Agnolo$^{3}$ \\
\vspace*{0.6cm}
\vspace*{0.6cm}
\it{\footnotesize $^1$ School of Natural Sciences, Institute for Advanced Study, Princeton, NJ 08540, USA} \\
\vskip5mm
\it{\footnotesize $^2$ \ Department of Physics and Astronomy, University of British Columbia, 6224 Agricultural Road, Vancouver, B.C. V6T 1Z1, Canada} \\
\vskip5mm
\it{\footnotesize $^3$ Theoretical Particle Physics Laboratory, Institute of Physics, EPFL, Lausanne, Switzerland}
\vskip5mm
\tt{danninos@ias.edu, tarek@phas.ubc.ca, raffaele.dagnolo@epfl.ch}



\end{center}
\vspace*{1.5cm}
\begin{abstract}
\noindent

Motivated by the near-horizon geometry of four-dimensional extremal black holes, we study a disordered quantum mechanical system invariant under a global $SU(2)$ symmetry. As in the Sachdev-Ye-Kitaev model, this system exhibits an approximate $SL(2,\mathbb{R})$ symmetry at low energies, but also allows for a continuous family of $SU(2)$ breaking marginal deformations. Beyond a certain critical value for the marginal coupling, the model exhibits a quantum phase transition from the gapless phase to a gapped one and we calculate the critical exponents of this transition. We also show that charged, rotating extremal black holes exhibit a transition when the angular velocity of the horizon is tuned to a certain critical value. Where possible we draw parallels between the disordered quantum mechanics and charged, rotating black holes.

\end{abstract}

\newpage
\setcounter{page}{1}
\pagenumbering{arabic}

\tableofcontents
\setcounter{tocdepth}{2}

\onehalfspacing


\section{Introduction}
Astrophysical black holes have been observed to be extremal within experimental errors, spinning at a rate remarkably close to the maximally allowed value set by their mass \cite{Reynolds:2013qqa, McClintock:2013vwa, Reynolds:2013rva}. A universal consequence of Einstein's equations is that such highly spinning objects develop a throat near their horizons, which becomes infinitely deep at saturation. Furthermore, apart from the axial $U(1)$ isometry, the geometry at extremality and deep within the throat becomes that of a two-dimensional anti-de Sitter space, where the Killing symmetries are enhanced to those of the conformal group in $(0+1)$-dimensions: $SL(2,\mathbb{R})$. The generic appearance of the AdS$_2$ symmetries in extremally rotating black hole geometries gives the holography of AdS$_2$ a distinguished role in the broader family of correspondences between anti-de Sitter space and conformal systems \cite{Strominger:1998yg,Sen:2011cn}.

We are therefore motivated to understand the holography of AdS$_2$ living near the horizon of highly spinning black holes. In order to do so, it will be convenient to consider a wider class of extremal black holes which are also electrically charged and appear as solutions of four-dimensional Einstein-Maxwell theory.
At fixed charge $Q$, Einstein-Maxwell theory admits a one-parameter family of $SL(2,\mathbb{R})\times U(1)$ invariant extremal geometries, discovered by Bardeen and Horowitz \cite{Bardeen:1999px}, and obtained as a near-horizon limit of the maximally rotating Kerr-Newman black hole. At a special point in parameter space, where the black hole no longer rotates, the $U(1)$ symmetry is enhanced to $SO(3)$ and the solution is the Bertotti-Robinson geometry: AdS$_2 \times S^2$ \cite{Bertotti:1959pf,Robinson:1959ev}.
These near-horizon geometries are obtained by taking a scaling limit of the original Kerr-Newman geometry, which involves rescaling the asymptotically flat time coordinate. This rescaling offsets the infinite redshift near the horizon, and what remains is the following solution to Einstein-Maxwell theory:
\begin{equation}\label{eq:bardhor}
ds^2 = \left(1-\frac{a^2 \sin^2\theta}{r_0^2}   \right) \left( - \frac{r^2}{r_0^2} dt^2 + \frac{r_0^2}{r^2} dr^2 + r_0^2 d\theta^2 \right) + \frac{r_0^2 \sin^2\theta}{ \left(1-\frac{a^2}{r_0^2}  \sin^2\theta \right)} \left( d\phi + \frac{2M a r}{r_0^4} dt \right)^2~,
\end{equation}
where $r_0^2 \equiv M^2 + a^2$ and $M^2 =Q^2 + a^2$. The relation between the ADM mass $M$ and electric charge $Q$ ensures that the horizon has vanishing temperature, but the parameter $a$ can take any value along the real axis. The horizon is located at $r=0$, so the above coordinates are in a co-rotating frame with respect to the horizon.
Summarizing, the geometry exhibits an $SL(2,\mathbb{R})\times U(1)$ isometry for all non-zero $a$, with the $U(1)$ being the axial symmetry.  When $a=0$ these isometries are enhanced to $SL(2,\mathbb{R})\times SO(3)$.

When $a$ is small we can view \eqref{eq:bardhor} as a deformation of the AdS$_2 \times S^2$ geometry, which gets corrected to leading order in $a$ by
\begin{equation}\label{eq:so3break}
\delta g_{\mu\nu}dx^\mu dx^\nu = \frac{ 4  a  r  \sin^2\theta}{Q} d\phi dt + \mathcal{O}(a^2)~.
\end{equation}
Notice that the above deformation is invariant under $t \to \lambda t$, $r \to r/\lambda$, but breaks the $SO(3)$ symmetry.
Our goal will be to model such an $SO(3)$ breaking deformation in a tractable large $N$ quantum system. The deformation (\ref{eq:so3break}) depends explicitly on the polar angle of the two-sphere, but not the azimuthal one. Thus, as mentioned, it breaks the original $SO(3)$ symmetry down to a $U(1)$ subgroup.

Holographically we might imagine a one-parameter family of $SL(2,\mathbb{R})$ invariant theories dual to the Bardeen-Horowitz geometries (recent work in this direction includes \cite{Bardeen:1999px,Guica:2008mu,Anninos:2008fx,Castro:2009jf,Detournay:2012pc}). We take here the view that the AdS$_2\times S^2$ point is captured, holographically, by some large $N$ quantum mechanical system with $SL(2,\mathbb{R})\times SU(2)$ symmetry.\footnote{We distinguish here $SO(3)$ from its double cover, $SU(2)$, since the building blocks of our quantum system will be fermions transforming under the fundamental of $SU(2)$. This is somewhat analogous to the $R$-symmetry group of $\mathcal{N}=4$ super Yang-Mills being $SU(4)$ and its holographic relation to the bulk $S^5$ with an $SO(6)\cong SU(4)$ isometry.} We might then view the deformation \eqref{eq:so3break} as turning on an $SU(2)$ breaking source in the original $SU(2)$ invariant quantum mechanics. Crucially, such a deformation must preserve the $SL(2,\mathbb{R})$.
In fact, from the black hole point of view, the coupling measuring the strength of the associated deformation is the angular velocity of the horizon, $\Omega$. This is the chemical potential for the angular momentum of the black hole. At finite electric charge, we will show that $\Omega$ cannot take arbitrarily large values for the Kerr-Newman geometry. Consequently, from the point of view of the dual quantum mechanics, we only require that the deformation remain marginal for some finite range of the coupling.\footnote{An example of marginal deformations that occur only for a finite range of the coupling at large $N$ is the case of a (classically marginal) $\phi^6$ deformation to the three-dimensional free $O(N)$ model. In this case, one can compute the $\beta$-function at large $N$ and show that it vanishes for small values of $\lambda_6$ \cite{Bardeen:1983rv}. However, for $\lambda_6$ above a certain value, the $\phi^2$ operator acquires a vacuum expectation value, indicating that the conformal symmetry is spontaneously broken.}

In this paper we take some simple steps in these directions. We study a disordered large $N$ quantum mechanics invariant under a global $SU(2)$ symmetry. This model resembles the Sachdev-Ye-Kitaev (SYK) model \cite{kitaev,Polchinski:2016xgd,Maldacena:2016hyu,Gross:2016kjj,Sachdev:2015efa,Anninos:2016szt,Jevicki:2016bwu,Anninos:2013nra} in that it is approximately conformal in the infrared, with low energy fluctuations controlled by the Schwartzian action. In addition, this system allows for an $SU(2)$-breaking deformation which we show is marginal for a finite range of couplings, mimicking some of the phenomenology of charged, rotating black holes.

\subsubsection*{Structure of the paper}

 We begin in section \ref{sec:KNtherm} by reviewing the thermodynamics of Kerr-Newman black holes and show that these black holes exhibit an interesting behavior as we approach a certain critical value of the angular velocity of the horizon. We then introduce a simple model of fermions with random masses in section \ref{sec:model}, which we proceed to solve to leading order in the large $N$ limit. The fermionic degrees of freedom are charged under a global $SU(2)$ symmetry. At low temperatures, the system exhibits an emergent local symmetry consisting of reparameterizations of time as well as a local $SU(2)$. These are broken to a global $SL(2,\mathbb{R}) \times SU(2)$ symmetry by the saddle point solution. In section \ref{sec:lowenergy} we discuss the soft modes associated with this breaking, whose action is induced by explicit breaking terms. We proceed to study the thermodynamics of this model in section \ref{sec:therm}. In section \ref{sec:breaking} we introduce a deformation that breaks the $SU(2)$ down to a $U(1)$ subgroup and establish that this deformation is marginal for a finite range of couplings, beyond which there is a transition to a gapped phase. In section \ref{sec:breakingtherm} we study the thermodynamics of the deformed model and show that the ungapped to gapped transiton is characterized by a non-analyticity in the partition function. We compare our model to the thermal properties of near-extremal charged and rotating black holes.
We end with a discussion on possible future directions in section \ref{sec:discussion}.

\section{Kerr-Newman thermodynamics}\label{sec:KNtherm}

In this section we review some of the the thermodynamic features of the Kerr-Newman solution, stressing the role of the angular velocity of the horizon. This quantity has an analog in the quantum mechanics models described in the next sections that controls the transition between different phases of the theory.

\subsection{Thermodynamic variables}

The general solution is labelled by the ADM mass $M$, the electric charge $Q$ and the ADM angular momentum $J$. These are conjugate to the temperature $T$, angular velocity $\Omega$ and electric potential $\Phi$. Depending on our ensemble, we can construct several different thermodynamic potentials. The one relevant to our discussion keeps $Q$, $T$ and $\Omega$ fixed. Equilibrium is achieved at the minimum of the following thermodynamic function:
\begin{equation}
\mathcal{G} (T,\Omega) = M -  T S  - \Omega J~,
\end{equation}
such that $\delta G = \left( \delta M - T \delta S - \Omega \delta J \right) =  0$, while keeping $Q$ fixed. A convenient parametrization of the thermodynamic potentials is obtained by trading $(M,Q,J)$ in favor of $(r_+,Q,a)$, where $r_+$ is the location of the black hole horizon:
\begin{equation}
r_+ \equiv M + \sqrt{M^2-a^2-Q^2}~.
\end{equation}
It follows that a regular solution only exists for $J< M \sqrt{M^2 - Q^2}$. In terms of these variables (in units where $G=\hbar = c = 1$):
\begin{equation}
S = {\pi} \, \left({r_+^2 + a^2} \right)~, \quad M = \frac{a^2+Q^2+r_+^2}{2 {r_+}}~, \quad J = a M ~.
\end{equation}
Thus, at equilibrium:
\begin{equation}
T = \frac{r_+ \left(1 - (a^2 + Q^2)/r_+^2 \right)}{4\pi\left(r_+^2 + a^2 \right)}~, \quad \Omega = \frac{a}{r_+^2 + a^2}~.
\end{equation}

Consider the rescaling:  $a \to Q a$, $r_+ \to Q r_+$, $M \to Q M$, $J \to Q^2 J$, $S \to Q^2 S$, $\Omega \to  \Omega/Q$ and $T \to T/Q$. Using this, we can set $Q=1$ in all equations, and reinstate it whenever necessary. Without loss of generality, we will always take $Q>0$.

\subsection{Extreme Kerr-Newman quantum roto-dynamics}
The extremal limit $T=0$ is obtained when:
\begin{equation}
r_+ = M = \sqrt{a^2+1}~,  \quad\quad \Omega = \frac{a}{(2a^2+1)} \in \left(-\frac{1}{2\sqrt{2}},\frac{1}{2\sqrt{2}} \right)~.
\end{equation}
Notice that, in contrast to when $Q=0$, at non-zero but fixed $Q$ the angular velocity of the horizon $\Omega$ is bounded from above and below. We can solve for $S(\Omega)$ and $J(\Omega)$ at $T=0$ obtaining:
\begin{equation}\label{jo}
S_\pm(\Omega)=\frac{2\pi}{1\pm s}~,\quad\quad {J_{\pm}(\Omega)} = \text{sign}(\Omega) \, \frac{\sqrt{(1\mp s) (3\pm s)}}{2 (1\pm s)}~,
\end{equation}
where we have defined:
\begin{equation}\label{eq:sdef}
s\equiv \sqrt{1-8\Omega^2}~.
\end{equation}
\begin{figure}
\begin{center}
\includegraphics[width=0.45\textwidth]{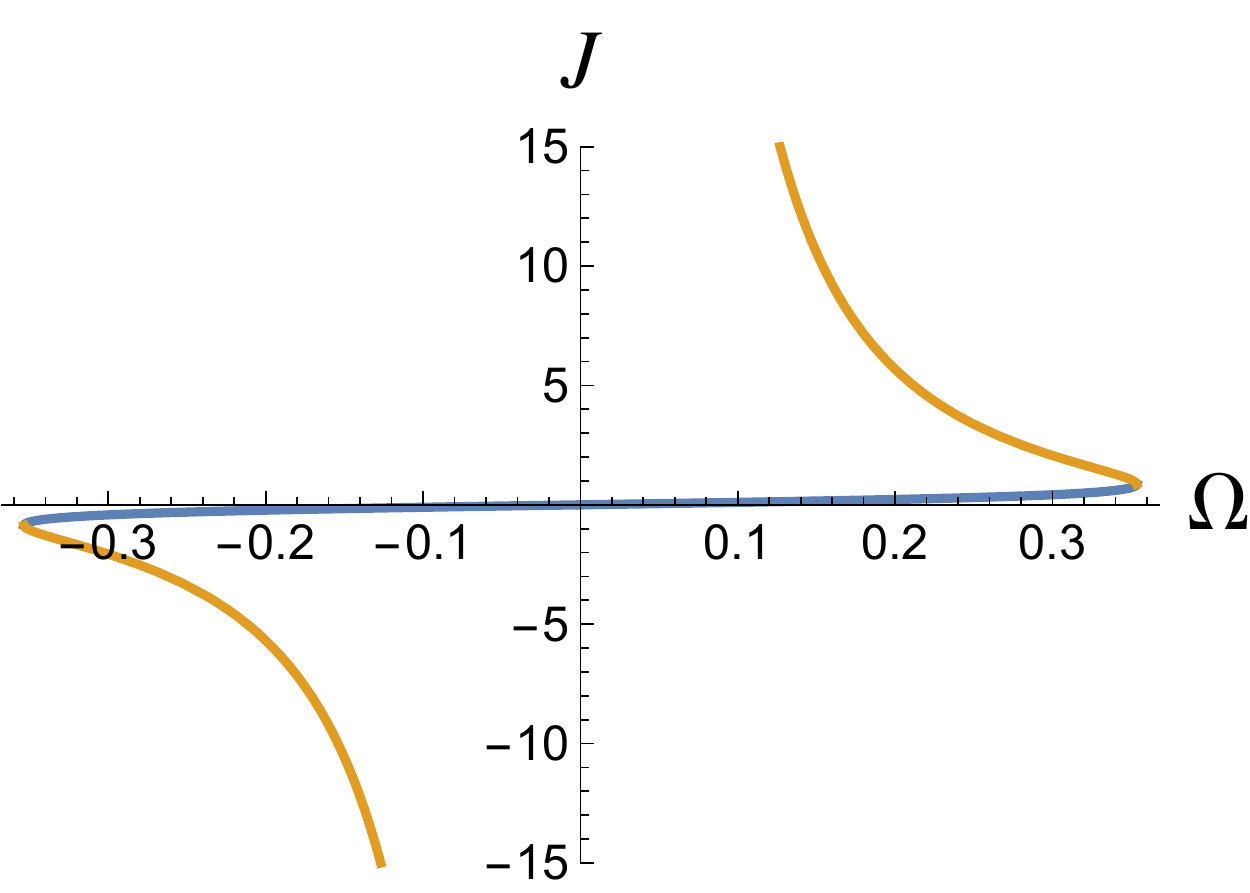}\hfill
\includegraphics[width=0.45\textwidth]{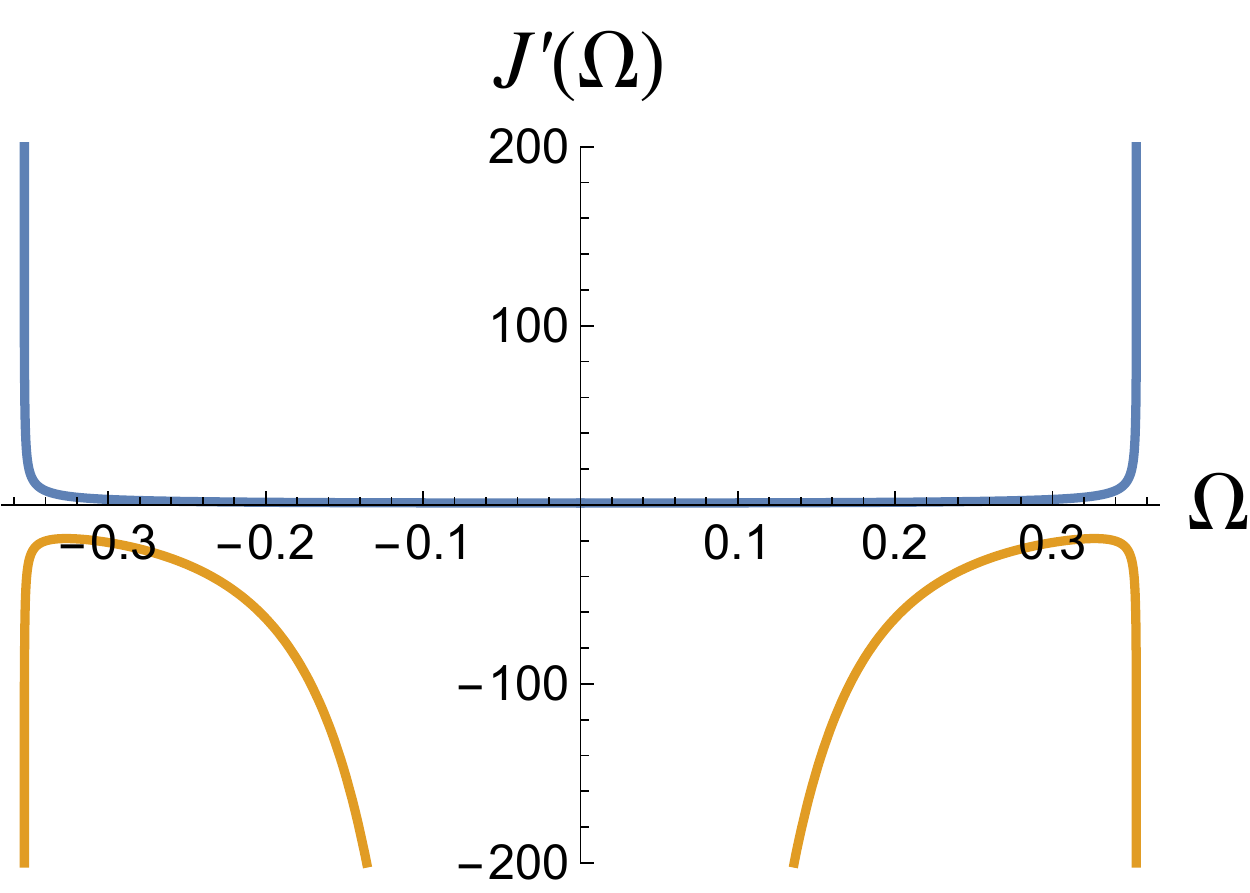}
\caption{\emph{Left}: $J(\Omega)$ vs. $\Omega$. The blue curve represents the $J_+(\Omega)$ branch and the orange curve the $J_-(\Omega)$ one. \emph{Right}: $J'(\Omega)$ displaying the divergent derivatives at $\Omega=\pm 1/\sqrt{8}$.}\label{jomega}
\end{center}
\end{figure}
The positive branch is continuously connected to the non-rotating extremally charge black hole. For small $\Omega$ ($s\approx 1$), the solution $J_+(\Omega) \approx \Omega$ and for $\Omega^2 > 1/8$ $J_+(\Omega)$ becomes complex. We present this in figure \ref{jomega}. Along the other branch, $J_-$ and $S_-$ both diverge at small $\Omega$. These branches meet at $\Omega=\pm1/\sqrt{8}$. Further note that:
\begin{equation}
\partial_{\Omega}  \left( J_\pm(\Omega) \right)^2 = 4 \sqrt{2} \, \text{sign}(\Omega) \frac{\sqrt{(1- s)(1+s)}}{(s\pm1)^3}\, \frac{1}{s}~.
\end{equation}
which diverges at $s = 0$, i.e. $\Omega^2 = 1/{8}$. Note that the angular momentum does not diverge at these values of $\Omega$, only its derivative does. Indeed, $\Omega(J)$ is extremized for these values of $J$ and therefore the derivative $\partial_J \Omega$ vanishes at these points. It follows that the inverse function must have corresponding derivatives that blow up. The extremal ADM mass is given by:
\begin{equation}
{M}^2_\pm = \frac{1}{2}\frac{(s\pm3)}{(s\pm1)}~.
\end{equation}
Notice that $M_+ = 1$ when $s=1$, which upon reinstating $Q$ is precisely the mass of an extremally charged Reissner-Nordstr{\"o}m black hole. It is also the case that $\partial_{\Omega} M_+$ diverges like $1/s$ in the limit $s \to 0$.

\subsection{Small temperature expansion of the free energy}

At finite temperature, it is difficult to write the free energy as a function of $\Omega$, $T$ and $Q$. However, we will mostly be interested in this function to leading order in a small temperature expansion. First it is convenient to express $a$ as a function of $r_+$ and $\Omega$ along the positive branch of the previous section:
\begin{equation}
a =  \frac{1-\sqrt{1-4 r_+^2 \Omega^2}}{2 \Omega }~.
\end{equation}
The temperature, as measured by an observer in the asymptotically flat region, can also be written as a function of $r_+$ and $\Omega$:
\begin{equation}\label{eq:temp}
T = \frac{2 r_+^2 \sqrt{1-4 r_+^2 \Omega^2}- \left(1+\sqrt{1-4 r_+^2 \Omega^2}\right)}{8 \pi  r_+^3}~.
\end{equation}
We can parameterize \eqref{eq:temp} by:
\begin{equation}
 r_+=\frac{1}{\sqrt{2}}\sqrt{\frac{s+3}{s+1}}+\delta r_+
 \end{equation}
where extremality occurs at $\delta r_+=0$. By inverting \eqref{eq:temp} order by order for $\delta r_+$ in a small $T$ expansion, we can thus express the free energy $\mathcal{G}$ at low temperatures:
\begin{equation}\label{GsmallT}
  \mathcal{G}=\frac{Q}{s+1}\left[\sqrt{\frac{(s+1)(s+3)^3}{32}}-2\pi Q \,T-\frac{(2\pi Q \,T)^2}{s\sqrt{2}}\sqrt{\frac{s+3}{s+1}}+\dots\right]~,
\end{equation}
where we have reintroduced $Q\neq 1$. The first term in (\ref{GsmallT}) is simply $(M_+-\Omega J_+)$ at exactly zero temperature. The term linear in $T$ has a coefficient given by the extremal entropy $S_+$ of the Kerr-Newman black hole. The term quadratic in $T$, measures certain thermodynamic fluctuations. Namely,
\begin{equation}
- \partial_T^2 \mathcal{G} = \frac{1}{T} C = Q^3 \, \frac{4\sqrt{2}\pi^2}{s(s+1)}\sqrt{\frac{s+3}{s+1}}~.
\end{equation}
This implies a positive specific heat linear in the temperature, characteristic of near extremal black holes. Finally, we note that the leading small temperature correction to the zero-temperature angular momentum (\ref{jo}) is given by:
\begin{equation}
\delta J_+ =  \text{sign}(\Omega) \, Q^3 \, T \,\frac{4\pi \sqrt{2}}{s}  \sqrt{\frac{1 - s}{ (1+s)^3}}~.
\end{equation}
Note that the sign of $\delta J_+$ is correlated with the sign of \eqref{jo} and grows in magnitude in the $s\to 0$ limit.

\subsection{Critical behavior}\label{BHPT}

Several of the thermodynamic quantities exhibit a non-analytic behavior at zero temperature in the limit $s\to0$.
For example the derivative of the angular momentum has a non-analytic divergence $\partial_\Omega J(\Omega) \sim 1/\sqrt{1-8 Q^2 \Omega^2}$. Viewing $\Omega$ as a tunable chemical potential in some putative dual large $N$ quantum system, the non-analytic behavior is suggestive of (quantum) critical behavior.\footnote{In the appendix we present a related non-analyticity of the zero-temperature partition function, expressed in terms of the Frolov-Thorne temperature \cite{Frolov:1989jh}.} We are going to observe this in our quantum mechanical models, where the role of $\Omega$ is played by the chemical potential of a global $SU(2)$ charge.

We can also consider non-analyticities in the $s \to 0$ limit at small but non-vanishing temperatures. This is subtle since a configuration with $s=0$ must also have exactly zero temperature. In other words, there is an order of limits that we need to specify --- either we first take the zero temperature limit or the $s\to0$ limit. In~(\ref{GsmallT}) we explicitly expanded in $T$ at fixed $s$.

From the near horizon point of view, we also need to specify the frame (and in particular the clock) that we pick to measure the temperature. Recall that when taking the near-horizon limit we had to redshift the clock in order to measure finite quantities. In view of this, we might imagine a rescaled near-horizon clock giving rise to a less singular low temperature expansion in the $s\to0$ limit. In the near extremal case, this ambiguity in the choice of clock is related to where we glue the near-horizon region to the asymptotically flat region. In the language of AdS$_2$ dilaton gravity \cite{Maldacena:2016upp}, this is the choice of the boundary value for the dilaton. If we rescale the clock by some factor $\lambda$, then $T \to T/\lambda$ and $\Omega \to \Omega/\lambda$. For $s$ to remain fixed we should also rescale $Q \to \lambda Q$. This has the effect of taking $\mathcal{G} \to \lambda \mathcal{G}$ and $C \to \lambda^2 C$. Picking $\lambda = s$, for example, leads to a non-singular $\mathcal{G}$ in the limit $s \to 0$, but we still have critical behavior at low temperatures $C \sim \sqrt{1-8Q^2\Omega^2}$.

As a final note, it would be interesting to understand what happens above the critical value $Q^2\Omega^2 = 1/8$. This `phase' is no longer described by an axisymmetric single-centered black hole, so if a solution exists, it may well be far more involved.
\newline\newline
In the next two sections we introduce and solve a quantum mechanical model that captures some of the qualitative features of Kerr-Newman thermodynamics. Parallels between the critical behavior described here and the phase structure of the quantum mechanical model are drawn in section~\ref{sec:breakingtherm}.

\section{Quenched disordered model}\label{sec:model}

To set the notation we first consider a supersymmetric quantum mechanics with a global $SU(2)$ $R$-symmetry group\footnote{This system is a simple toy model for the kind of systems related to the low energy open string sector of wrapped intersecting branes \cite{Denef:2002ru}. However, none of what follows relies on the string theoretic interpretation of this system and we will keep the discussion general. Generalizations of Sachdev-Ye-Kitaev systems with global symmetries were considered in \cite{Gross:2016kjj,Yoon:2017nig}.} that was already studied in~\cite{Anninos:2013nra, Anninos:2016szt}. In section~\ref{sec:breaking} we break this $R$-symmetry by introducing a deformation, in order to make contact with rotating black holes.

The field content is given by $N$ chiral supermultiplets $\{ \phi_\alpha, \psi^a_\alpha, F_\alpha\}$ containing a complex scalar $\phi_\alpha$,  a two-component spinor $\psi^a_\alpha$, and an auxiliary complex scalar $F_\alpha$. The index $a=1,2$ is a fundamental $SU(2)$ index, and $\alpha = 1,\ldots,N$. The model has a quadratic superpotential:
\begin{equation}
W(\phi)  =  \Gamma_{\alpha\beta} \phi_\alpha \phi_\beta~,
\end{equation}
where the $\Gamma_{\alpha\beta}$ are $N\times N$ complex numbers which we take to be distributed randomly, following a zero-mean Gaussian distribution:
\begin{equation}
P(\Gamma_{\alpha\beta}) = \frac{N}{\pi \gamma^2} {e^{-N\Gamma_{\alpha\beta}\bar{\Gamma}_{\alpha\beta}/ \gamma^{2}}}~.
\end{equation}
The Euclidean action for this superpotential reads:
\begin{equation}\label{eq:origaction}
S_E = \int d\tau \left[ \bar{\psi}^{\dot{a}}_\alpha \dot{\psi}^a_\alpha + \dot{\bar{\phi}}_\alpha \dot{\phi}_\alpha  -  \bar{F}_\alpha F_{\alpha} + \left(\Gamma_{\alpha\beta} ( F_\alpha \phi_\beta+F_\beta \phi_\alpha) + \Gamma_{\alpha\beta} \, \psi_\alpha^a \epsilon_{ab} \psi_\beta^b + h.c. \right)\right]~,
\end{equation}
where $\epsilon_{ab}=-i\boldsymbol{\sigma}_y=-\epsilon_{\dot{a}\dot{b}}$ is the Levi-Civita symbol. The $SU(2)$ symmetry acts only on the fermions:
\begin{equation}
\psi^a \to {U^{a}}_{b} \psi^b~,  \quad \bar{\psi}^{\dot{a}} = \bar{\psi}^{\dot{b}} {{U^{\dag}}_{\dot{b}}}^{\dot{a}} \, ~,
\end{equation}
where
\begin{equation}
{U^{a}}_{b}  = \begin{pmatrix}e^{-i(\xi+\phi)}\cos\theta &-e^{-i(\xi-\phi)}\sin\theta\\
~~e^{i(\xi-\phi)}\sin\theta &~~~e^{i(\xi+\phi)}\cos\theta
\end{pmatrix}
\end{equation}
is a group element of $SU(2)$ in the fundamental representation. Consequently, the fermions will play the main role in the physics that we discuss. It is worth emphasizing that the $U(1)$ fermion number generator is not conserved in this model due to the quadratic couplings in the action, and therefore our action is not $U(2)$ invariant.  In Fourier space we have
\begin{multline}
S_E = \int \frac{d\omega}{2\pi} \left[i\omega\, \bar{\psi}^{\dot{a}}_\alpha(\omega) {\psi}^a_\alpha(-\omega) + \omega^2\, {\bar{{\phi}}}_\alpha(\omega) {\phi}_\alpha(-\omega)  -  \bar{F}_\alpha(\omega) F_{\alpha}(-\omega) + \right. \\ \left.  \left(\Gamma_{\alpha\beta} (F_\alpha(\omega) \phi_\beta(-\omega) +F_\beta(\omega) \phi_\alpha(-\omega))+ \Gamma_{\alpha\beta} \, \psi_\alpha^a(\omega) \epsilon_{ab} \psi_\beta^b(-\omega) + h.c. \right)\right]~.
\end{multline}
Since we assume the $\Gamma_{\alpha\beta}$ are drawn from a random Gaussian ensemble with variance $\gamma^2/N$, at large $N$ we can work in the disorder-averaged theory. Along the lines of \cite{kitaev,Polchinski:2016xgd,Maldacena:2016hyu,Gross:2016kjj,Sachdev:2015efa,Anninos:2016szt,Jevicki:2016bwu,Anninos:2013nra}, we take the average using the replica trick. This entails introducing $n$ copies of the system and exploiting the simple formula,
\begin{equation}
\log Z_{\Gamma_{\alpha\beta}}[\beta] = \lim_{n\to 0} \partial_n \left( Z_{\Gamma_{\alpha\beta}}[\beta]\right)^n~,
\end{equation}
such that the disorder averaging can be performed on $Z_{\Gamma_{\alpha\beta}}^n$ as opposed to $\log Z_{\Gamma_{\alpha\beta}}$. We use a replica symmetric ansatz in what follows, since it was shown in \cite{Anninos:2016szt}, that this is the dominant saddle. This ansatz implies that the number of replicas, $n$, does not play a crucial role and simply appears as an overall factor of the on-shell action.

Notice that the scalars and fermions are completely decoupled in (\ref{eq:origaction}), and hence any mixed correlator between them vanishes on shell. Furthermore, the $SU(2)$ invariance guarantees that $\langle \bar{\psi}^{\dot{a}}_\alpha(\tau)~\psi^b_\alpha(\tau') \rangle~$ for $a\neq b$ also vanish on-shell. Thus we need only introduce the following bi-local variables:
\begin{equation}
Q(\tau,\tau') = \frac{1}{N} \langle \bar{\phi}_\alpha(\tau) \phi_\alpha(\tau') \rangle~, \quad\quad S^a(\tau,\tau') = \frac{1}{N} \langle \bar{\psi}^{\dot{a}}_\alpha(\tau)~\psi^a_\alpha(\tau') \rangle~,
\end{equation}
in terms of which the disorder averaged effective action is found to be:
\begin{multline}\label{eq:effaction}
\frac{S_{\rm eff}}{N n} = \int {d\tau}d\tau' \left[ \left(- \partial^2_\tau Q(\tau,\tau') +\sum_a \partial_{\tau'} S^a(\tau,\tau')  \right) \delta(\tau-\tau') - \gamma^2 S^1(\tau,\tau')S^2(\tau,\tau') \right]  \\ +  \text{tr} \log(1+\gamma^2 Q(\tau,\tau')) -\text{tr} \log Q(\tau,\tau') +\sum_a\text{tr} \log S^a(\tau,\tau')~.
\end{multline}
In Fourier space this reads
\begin{multline}\label{eq:effactionFour}
\frac{S_{\rm eff}}{N n} = \int \frac{d\omega}{2\pi} \Big[ \omega^2 Q(\omega) + \log(1+\gamma^2 Q(\omega)) - \log Q(\omega)   \\    +\sum_a\left( i \omega S^a(\omega)  + \log S^a(\omega)\right) - \gamma^2 S^1(\omega)S^2(-\omega) \Big]~,
\end{multline}
where
\begin{equation}
Q(\omega) = \frac{1}{N} \langle \bar{\phi}_\alpha(\omega) \phi_\alpha(-\omega) \rangle~, \quad\quad S^a(\omega) = \frac{1}{N} \langle \bar{\psi}^{\dot{a}}_\alpha(\omega)~\psi^a_\alpha(-\omega) \rangle~.
\end{equation}
To derive the above action we have assumed time translation invariance, $Q(\tau,\tau') = Q(\tau-\tau')$ and $S^a(\tau,\tau') = S^a(\tau-\tau')$. Since certain correlators are guaranteed to vanish on-shell we have set them to zero in (\ref{eq:effaction}-\ref{eq:effactionFour}). These streamlined expressions are obtained at the expense of having a bi-local effective action \eqref{eq:effaction} which is not manifestly $SU(2)$ invariant.

\subsection{Saddle point solutions}

\begin{figure}
\begin{center}
\includegraphics[width=0.7\textwidth]{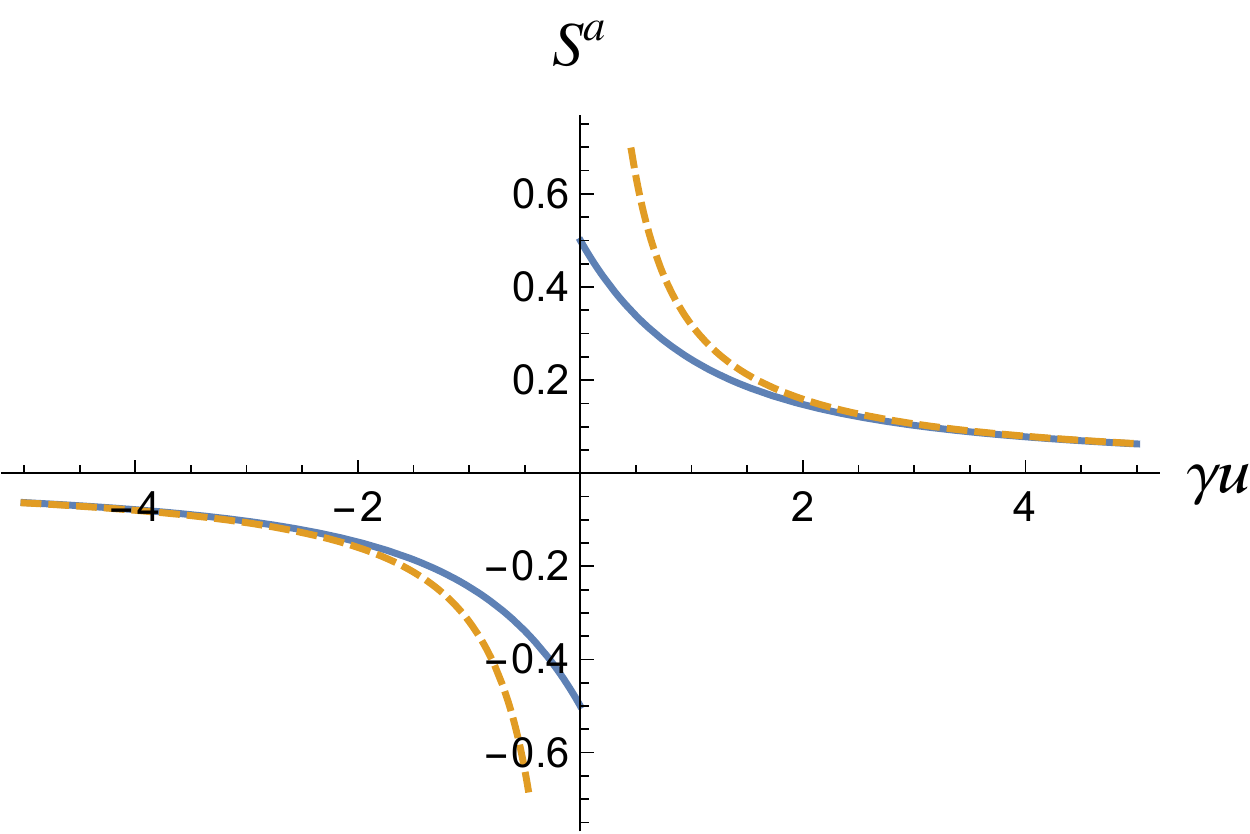}
\caption{The solid blue curve is the exact result for $S^a(u)$, while the dashed orange curve is the large time separation approximation $S_{\rm low}^a(u)= 1/\pi \gamma u$.}\label{fig:struve}
\end{center}
\end{figure}

The original action (\ref{eq:effactionFour}) has a CT (complex conjugation and time reversal) invariance  which acts on the fields as
\begin{equation}\label{eq:CT}
\omega \to -\omega~, \quad\quad \bar{\psi}_\alpha^{\dot{a}}(\omega) \to \psi_\alpha^{a}(-\omega)~,\quad\quad \phi_\alpha(\omega)\to \bar{\phi}_\alpha(-\omega)~, \quad\quad \Gamma_{{\alpha\beta}} \to \bar{\Gamma}_{{\alpha\beta}}~.
\end{equation}
So $S^a(\omega) = -S^a(-\omega) = \overline{S^a(-\omega)}$ for CT invariant states. The saddle point equations for $S^a$ and $Q$ are found to be
\begin{eqnarray}
\frac{1}{S^1(\omega)} &=& - i \omega   + \gamma^2 S^2(-\omega)~, \label{eq:s1}\\
\frac{1}{S^2(\omega)} &=& - i \omega   + \gamma^2 S^1(-\omega)~, \label{eq:s2}\\
\frac{1}{Q(\omega)} &=&  \omega^2  + \frac{\gamma^2}{1+ \gamma^2 Q(\omega)} ~,
\end{eqnarray}
and the solutions are
\begin{equation}\label{solns}
  Q(\omega) = \frac{2 \, \omega^{-2}}{1+\sqrt{1+\left(\frac{\omega}{2\gamma}\right)^{-2}}}~, \quad S^a(\omega) = \frac{i}{\gamma}\left(-\frac{\omega}{2\gamma}+\text{sign}(\omega)\sqrt{1+\left(\frac{\omega}{2\gamma}\right)^{2}}\right)~.
\end{equation}
Note that they obey the supersymmetric Ward identity $S(\omega)=  i \omega Q(\omega)$. At high frequencies, they reproduce the free correlators of a scalar and a fermion. At low frequencies, they become
\begin{equation}\label{eq:lowfreq}
Q_{\rm low}(\omega) = \frac{1}{\gamma|\omega|}~, \quad\quad S_{\rm low}^a(\omega) = i \frac{\text{sign}(\omega)}{\gamma}~.
\end{equation}
These are conformally invariant two-point functions. We can Fourier transform (\ref{solns}) back to Euclidean time and obtain the exact correlator:
\begin{equation}
S^a(u)  = \frac{I_1(2 |u | \gamma)-\pmb{L}_1(2 |u | \gamma)}{2 u \gamma}~, \quad\quad u \equiv \tau-\tau'~,
\end{equation}
where $\pmb{L}_1(z)$ is the modified Struve function and $I_1(z)$ is the modified Bessel function. At large time separations this simplifies to
\begin{equation}
S_{\rm low}^a(u) = \frac{1}{\pi \gamma u}~.
\end{equation}
Thus, at low energies, we observe that the correlation functions exhibit an emergent scale invariance, as shown in figure \ref{fig:struve}.
Using the integral representation of the Struve and Bessel functions we can write:
\begin{equation}
 \frac{I_1(2 |u | \gamma)-\pmb{L}_1(2 |u | \gamma)}{2 u \gamma} =\begin{cases} ~~\frac{1}{\pi \gamma} \, \int_0^{2\gamma} dE \sqrt{ 1- \left( \frac{E}{2\gamma}\right)^2 } \, e^{-E u}~, &u>0\\
 -\frac{1}{\pi \gamma} \, \int_0^{2\gamma} dE \sqrt{ 1- \left( \frac{E}{2\gamma}\right)^2 } \, e^{E u}~, &u<0
\end{cases}
\end{equation}
From this expression we can read off the spectral density (see e.g. \cite{sachdev2011} or appendix A of \cite{Davison:2016ngz}):
\begin{equation}
\rho^1(E) =\rho^2(E)= \frac{1}{\gamma}  \sqrt{ 1- \left( \frac{E}{2\gamma}\right)^2 }~, \quad\quad\quad |E| < 2\gamma~,
\end{equation}
which vanishes for $|E| > 2\gamma$. The spectral density can also be obtained from \eqref{solns} via:
\begin{equation}\label{eq:specdens}
  \rho^a(E)=\text{Im}\,S^a\left(-i\left(E+i0^+\right)\right)~,
\end{equation}
which requires us to define $\text{sign}(\omega)$ for complex frequencies, which we do by taking $\text{sign}(\omega)\rightarrow \omega/\sqrt{\omega^2}$.
The spectral density is non-negative, as required by unitarity \cite{sachdev2011}. Furthermore $\rho(E)$ is a semi-circle, reminiscent of Wigner's semi-circle for the eigenvalues of random Gaussian matrices. This stems from the fact that our model is quadratic in the fermions, with the mass matrix drawn from a random Gaussian ensemble.

We would also like to emphasize that the $SU(2)$ generators, $\hat{\bold{J}} = \bar{\psi}^{\dot{a}} \boldsymbol{\sigma}^{\dot{a}a} \psi^{a}$, have vanishing expectation values in this state. We think of this as analogous to the fact that AdS$_2\times S^2$ has an $SO(3) \cong SU(2)$ isometry. In fact, correlators of increasingly complicated operators charged under $SU(2)$ decay increasingly fast at late times. This is similar to the no hair theorem causing the redshift of any features on the two-sphere horizon of AdS$_2 \times S^2$.

\subsection{Low energy effective action}\label{sec:lowenergy}

The emergence of an $SL(2,\mathbb{R})$ implies that at low frequencies, or equivalently large time separations, the action exhibits an enhanced group of symmetries \cite{Sachdev:1992fk,parcollet,kitaev}:
\begin{equation}
\tau \to f(\tau)~, \quad S^{a}(\tau,\tau') \to  f'(\tau)^{1/2} S^a(f(\tau),f(\tau'))  f'(\tau')^{1/2}~, \quad Q(\tau,\tau') \to Q(f(\tau),f(\tau'))~.
\end{equation}
The above constitute the group of diffeomorphisms mapping the line to itself. The saddle point solutions spontaneously break this group to an $SL(2,\mathbb{R})$ subgroup given by:
\begin{equation}
\tau \to \frac{a\tau+b}{c\tau+d}~, \quad\quad a d - b c = 1~.
\end{equation}
The scaling dimension of $\phi$ is $\Delta_\phi = 0$, while that of $\psi$ is $\Delta_\psi = 1/2$.\footnote{The zero mode of $\phi$ is divergent and must be treated with care. One way to do so is to consider that $\phi$ is a coordinate in some compact space. This is what happens in more elaborate versions of the above model \cite{Denef:2002ru}.} In addition to the one-dimensional diffeomorphism group, the low frequency action exhibits an additional emergent local symmetry.\footnote{This is not apparent in the way we have written the action in (\ref{eq:effaction}). This is because we have used the fact that certain correlators vanish on shell. When these functions are reintroduced into the disorder averaged theory, the $SU(2)$ invariance becomes apparent.} This is an $SU(2)$ gauge symmetry acting as:
\begin{equation}\label{eq:su2gauge}
  S^a\equiv S^{\dot{a}a}(\tau,\tau') \to {{U^{\dag}}_{\dot{b}}}^{\dot{a}}(\tau)  S^{\dot{b}b}(\tau,\tau')   {U^{a}}_{b}(\tau')~, \quad\quad \forall \; \tau,\tau'~.
\end{equation}
Again, $Q(\tau,\tau')$ does not transform under this symmetry. The reason the $SU(2)$ symmetry is enhanced to a local symmetry is that in the low frequency limit we drop the kinetic terms of the fermions. However, in a small derivative expansion, we will induce an effective action for the $SU(2)$ gauge field.
We now proceed to derive the low energy effective action describing the time-reparametrization and local $SU(2)$ rotation modes.

We begin with the action for the fermions:
\begin{equation}\label{action}
\frac{S_{\rm eff}}{Nn}=\sum_a\text{tr}\log S^a+\int d\tau d\tau^\prime\left[ \delta(\tau-\tau^\prime)\sum_a\partial_{\tau'} S^a-\gamma^2 S^1S^2\right]\, .
\end{equation}
The action and equations of motion:
\begin{equation}\label{eom}
-\partial_{\tau'}\,S^a(\tau,\tau')-\gamma^2\int du\,S^b(\tau,u)\,S^a(u,\tau')=-\delta(\tau,\tau')
\end{equation}
are approximately invariant under the following symmetry:
\begin{equation}\label{symmetry}
\tau\rightarrow f(\tau)~,\quad\quad\quad\quad S^a(\tau,\tau')\rightarrow f'(\tau)^{1/2}S^a(f(\tau),f(\tau'))f'(\tau')^{1/2}
\end{equation}
if we drop the derivative terms in (\ref{action}) and (\ref{eom}). Without the derivative term, (\ref{eom}) admits the following solution:
\begin{equation}\label{eq:lowenergysaddle}
S^a(\tau,\tau')=\frac{1}{\pi\gamma}\frac{1}{\tau-\tau'}
\end{equation}
which becomes
\begin{equation}\label{generalsol}
S^a(\tau,\tau')=\frac{1}{\pi\gamma}\frac{f'(\tau)^{1/2}f'(\tau')^{1/2}}{f(\tau)-f(\tau')}
\end{equation}
after transforming by the symmetry (\ref{symmetry}).

Since (\ref{symmetry}) is not an exact symmetry of the Lagrangian it should contribute to the action, as should the approximately local $SU(2)$ symmetry. It can be verified, for example via an explicit coset construction~\cite{Coleman:1969sm,Callan:1969sn,Volkov:1973vd, Ivanov:1975zq,Low:2001bw}, that in the absence of explicit breaking these massless modes have zero action. So we are left with evaluating the contribution from the explicit breaking of these symmetries. This arises from the kinetic term:
\begin{multline}
  \frac{S^{\rm breaking}_{\rm eff}}{Nn}=\sum_{a,b=1}^2\int d\tau d\tau^\prime \delta(\tau-\tau^\prime)\,\partial_{\tau'}\left\{\left({{U^\dag}_{\dot{b}}}^{\dot{a}}(\tau){U^a}_{b}(\tau')\right){f'(\tau)^{1/2}f'(\tau')^{1/2}} \,S^{\dot{b}b}(f(\tau),f(\tau'))\right\}~,
\end{multline}
with $S^{\dot{a}a}(\tau,\tau')$ given in (\ref{eq:lowenergysaddle}). The above action displays a UV divergence at $\tau=\tau'$. The divergence is an artifact of taking the low energy approximation, since the full system is UV finite. We propose to regularize this by performing in the following simple deformation $\delta(\tau-\tau')\rightarrow\delta(\tau-\tau^\prime-\epsilon)$ and compute the action order by order in $\epsilon$. This leads to
\begin{equation}
  \frac{S^{\rm breaking}_{E}}{Nn}
  =\frac{2}{\pi\gamma}\int d\tau\left(\frac{1}{\epsilon^2}-\frac{1}{12}\lbrace f(\tau),\tau\rbrace+\frac{1}{4}\text{Tr}\left[\partial_\tau U^\dagger\cdot\partial_\tau U\right]+O\left(\epsilon^2\right)\right)~,\label{eq:schwarzaction}
\end{equation}
where
$\lbrace f(\tau),\tau\rbrace$ is the expected Schwarzian derivative \cite{Maldacena:2016hyu}:
\begin{equation}
  \lbrace f(\tau),\tau\rbrace=\frac{f'''}{f'}-\frac{3}{2}\left(\frac{f''}{f'}\right)^2~.
\end{equation}
The additional term, involving the unitary matrices, is that of a particle moving in an $SU(2)$ group manifold. It describes the low energy fluctuations of the $SU(2)$ charged sector \cite{Anninos:2015eji}.

\section{Low temperature thermodynamics of the $SU(2)$ invariant model}\label{sec:therm}
Having solved for the correlation functions of this quantum mechanical system, we now have all the ingredients necessary to study its thermodynamics. In what follows we only focus on the thermodynamics of the fermionic fields $\psi_\alpha^a(\tau)$ and ignore the contributions from the bosonic fields $\phi_\alpha(\tau)$. Connections between the thermodynamics of this model and those of charged rotating black holes will be made in later sections.


At low temperatures, the free energy admits an expansion:
\begin{equation}
\frac{1}{Nn}\log Z=- \frac{\beta F}{N n}=-\beta E_0+S_0+\frac{C(\beta)}{2}+\dots
\end{equation}
where $E_0$ is the ground state energy and $S_0$ is the ground state entropy (both divided by $Nn$) and $C(\beta)$ is a function whose dependence on the temperature is a fixed negative power of $\beta$.
The expectation value of the energy is related to derivatives of the free energy with respect to $\beta$:
\begin{equation}
\left\langle E\right\rangle=\frac{1}{Nn}\partial_\beta(\beta F)
\end{equation}
 which allows us to relate $C(\beta)$ to the low temperature contribution to the total energy (above the ground state energy $E_0$):
\begin{equation}
E_{\rm low}\equiv- \frac{1}{2} \partial_\beta {C(\beta)}~.
\end{equation}
Given our saddle point approximation, we can compute thermodynamic quantities from the on shell action, which, at finite temperature can be written as a sum over discrete frequencies $\omega_n = 2\pi\left(n+1/2\right)/\beta$ with $n\in \mathbb{Z}$. Let us begin by computing the energy of the system $\langle E\rangle=E_0+E_{\rm low}+\dots$. To do so we can take a $\beta$ derivative of the on-shell action where $S^a(\omega_n)$ take their on-shell values (\ref{solns}). This calculation is simplified by the fact that we can use the on-shell conditions (\ref{eq:s1}-\ref{eq:s2}), yielding:
\begin{equation}\label{eq:eexpect}
\langle E\rangle=-\frac{2}{\beta} \sum_{l\in \mathbb{Z}} \left[\frac{2\pi i (l+{1}/{2})}{\beta} S^1(\omega_l) + 1 \right]~.
\end{equation}
 We have not been able to perform the above sum analytically, however we can still obtain analytic expressions for $E_0$ and $E_{\rm low}$.

To calculate $E_{\rm low}$ we need not use the exact expressions for the correlators. Instead, we can extract it from the approximate low frequency solutions (\ref{eq:lowfreq}) as follows:
\begin{equation}\label{Zzero}
E_{\rm low}=- \frac{2}{\beta} \sum_{l\in \mathbb{Z}} \left[\frac{2\pi i (l+{1}/{2})}{\beta} S_{\rm low}^1(\omega_l) + 1 \right]~.
\end{equation}
Inserting the approximate correlator into (\ref{Zzero}) gives a divergent sum, but this is of course an artifact of our low energy approximation. To obtain an analytic expression, we use a $\zeta$-function regularization scheme, namely
\begin{equation}\label{zetareg}
\sum_{l = 0}^\infty \left(l+\frac{1}{2}\right) \equiv \zeta(-1,1/2) =  \frac{1}{24}~, \quad\quad \sum_{l = 1}^\infty 1 \equiv \zeta(0) =  -\frac{1}{2}~.
\end{equation}
Using this regularization, we find the following low temperature approximation:
\begin{equation}\label{zetaE}
E_{\rm low} = \frac{\pi}{3\beta^2 \gamma}~.
\end{equation}
The function $C=\partial_T E_{\rm low}$ can be identified with the specific heat, which is linear in $T$.

Equivalently, we can also compute $E_{\rm low}$ or $C(\beta)$ from the Schwarzian effective action (\ref{eq:schwarzaction}), using a map from the line to the Euclidean circle. To do this, we notice that at low temperatures the free energy contains a term \cite{Maldacena:2016hyu}
\begin{equation}
  -\frac{\beta F}{N n}\supset +\frac{1}{6\pi\gamma}\int_0^\beta d\tau \left\lbrace \tan\left(\frac{\pi\,\tau}{\beta}\right),\tau\right\rbrace=\frac{\pi}{3\beta\gamma}~,
\end{equation}
giving $C={2\pi}/{3\beta \gamma}$ as well as (\ref{zetaE}) for $E_{\rm low}$.

\begin{figure}
\begin{center}
\includegraphics[width=0.49\textwidth]{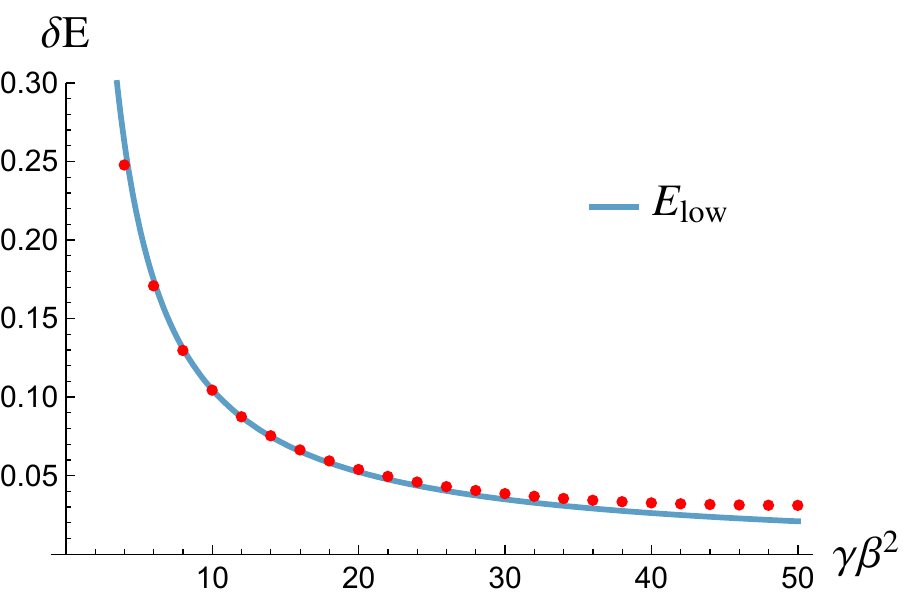}
\includegraphics[width=0.49\textwidth]{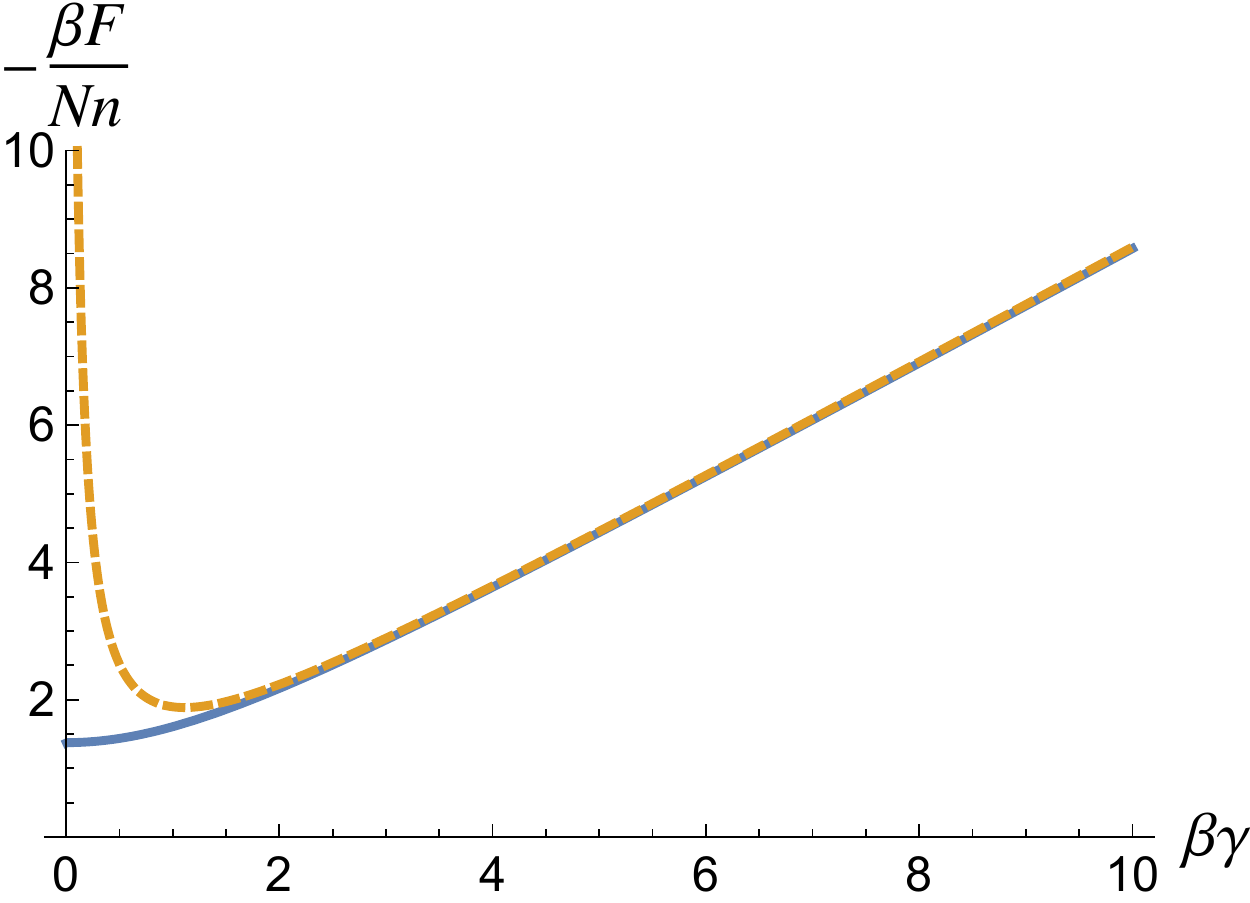}
\end{center}
\caption{\emph{Left:} plot of $\delta E(\beta) = \langle E(\beta)\rangle-E_0$ and the $\zeta$-regularized value $E_{\rm low}$. We have performed the sum in (\ref{eq:eexpect}) numerically using the exact expression for $S^a(\omega_n)$ and  cutoff our sum above and below at $n_c=\pm2.5\times 10^3$. As $n_c$ increases the plots become increasingly close. \emph{Right}: Comparison between numerically computed $-\beta F/Nn$ obtained by evaluating the on-shell action (solid curve) and the low temperature approximation $-\beta E_0+C(\beta)/2$ (dashed curve). At low temperatures these curves align implying that the zero temperature entropy $S_0$ vanishes.}\label{fig:numcompare}
\end{figure}

To compute the ground state energy $E_0$ we repeat the calculation at exactly zero temperature. This yields
\begin{equation}
E_0 = -2\times \int_{\mathbb{R}} \frac{d\omega}{2\pi} \left( i\omega\,S^1(\omega)+1\right) = -\frac{8 \gamma}{3\pi}~,
\end{equation}
where $S^1(\omega)$ is the exact solution given in (\ref{solns}). We can check our analytic expressions against the numerically evaluated energy $\langle E\rangle$ given in (\ref{eq:eexpect}). In figure \ref{fig:numcompare} we show a plot $\langle E\rangle -E_0$ as compared with $E_{\rm low}$ obtained via $\zeta$-function regularization in (\ref{zetaE}). These are in excellent agreement.

To extract the zero temperature entropy $S_0$, we must compute the finite temperature on-shell action for the fermionic degrees of freedom numerically. This requires us to deal with the UV divergence at high frequency, which we do by ensuring that the entropy at high temperatures is simply the dimension of the Hilbert space $2 Nn \log 2$. We plot the result in the right panel of figure \ref{fig:numcompare}. Comparing it with $-\beta E_0+C(\beta)/2$ at low temperatures shows that $S_0$ vanishes for our model.

Combining everything, the low temperature partition function can be written as:
\begin{equation}
-\frac{\beta F}{Nn}=\frac{1}{Nn}\log Z[\beta] =  \frac{8\beta\gamma}{3\pi}  +  \frac{\pi}{3\beta \gamma}+  \ldots
\end{equation}
The system has a specific heat linear in the temperature at low temperatures, $C={2\pi T}/{3 \gamma}$. This is a characteristic behavior of near-extremal black holes. A system with exact conformal invariance cannot have a specific heat linear in the temperature, since there is no scale in the problem. On the other hand, a linear specific heat at low temperatures implies a large density of states all the way down to zero energy, implying the presence of certain soft modes. As we saw, the modes giving rise to this behavior correspond to low energy modes appearing due to the breaking of the diffeomorphism group.



\section{$SU(2)$ breaking marginal deformation}\label{sec:breaking}

We now consider the following deformation to the original $SU(2)$-invariant action (\ref{eq:origaction}):
\begin{equation}\label{defS}
S_E\rightarrow S_E- z \int d\tau\, \bar{\psi}^{\dot{a}}_\alpha(\tau) \boldsymbol{\sigma}^{\dot{a}b}_z \psi^b_\alpha(\tau)~,
\end{equation}
which corresponds to turning on a chemical potential $z$ for the angular momentum operator\footnote{Here and in the following we call the charges associated with our internal $SU(2)$ symmetry ``angular momentum" given the analogy with black hole physics.} 
in the $\hat{z}$ direction $\hat{J}_z\equiv\bar{\psi}^{\dot{a}}_\alpha(\tau) \boldsymbol{\sigma}^{\dot{a}b}_z \psi^b_\alpha(\tau)$. The above operator corresponds to deforming the disorder-averaged theory (\ref{eq:effactionFour}) in the following way:
\begin{equation}
  \frac{S_{\rm eff}}{Nn}\rightarrow \frac{S_{\rm eff}}{Nn}-z\int \frac{d\omega}{2\pi}\,\left(S^1(\omega)-S^2(\omega)\right)~.
\end{equation}
Marginal operators in $(0+1)$-dimensional fixed points have $\Delta=1$. In the undeformed theory, the scaling dimension of the fermions at the infrared fixed point is $\Delta_\psi=1/2$. So the deformation (\ref{defS}) may describe a marginal deformation of the infrared fixed point\footnote{Relevant deformations of the SYK model were considered in \cite{Banerjee:2016ncu,Anninos:2017hhn}.}. The deformation explicitly breaks the original $SU(2)$ symmetry to a $U(1) \subset SU(2)$. Notice that the engineering dimension of $z$ has units of temperature, such that $z \beta$ is a dimensionless quantity, as is $\gamma \beta$. So we can work in units where $\beta=1$ (unless otherwise specified) whereby the low temperature limit corresponds to taking large $\gamma$ and $z$ with $\gamma/z$ fixed.

With the addition of the deformation (\ref{defS}), the action is invariant under:
\begin{equation}
\omega \to -\omega~, \quad\quad \bar{\psi}_\alpha^i (\omega) \to \psi_\alpha^i (-\omega)~, \quad\quad z \to -z~,\quad\quad\Gamma_{\alpha\beta}\to\bar{\Gamma}_{\alpha\beta}~.
\end{equation}
This indicates that to restore the CT invariance described in~(\ref{eq:CT}) we supplement it with an additional $z\rightarrow-z$ flip. This symmetry implies that the on-shell correlators respect: $S^1(\omega,z) = -{S^1(-\omega,-z)}$ and ${S^1(-\omega,z)} = \overline{S^1(\omega,z)}$.

\subsection{Deformed saddle point solutions}

As mentioned, given that $\Delta_\psi = 1/2$ in the IR, the deformation (\ref{defS}) might be marginal at low energies, since it is built out of two fermions. Let us verify that this is indeed the case. The disorder-averaged equations are now given by
\begin{eqnarray}
\frac{1}{S^1(\omega,z)} &=& - i \omega + z  + \gamma^2 S^2(-\omega,z)~, \\
\frac{1}{S^2(\omega,z)} &=& - i \omega - z  + \gamma^2 S^1(-\omega,z)~.
\end{eqnarray}
Let us first assume that the deformation is small, i.e. $|z|<2\gamma$. In this regime the exact correlator for the fermions is given by
\begin{equation}\label{eq:exactzdef}
  S^1(\omega,z) = \frac{1}{\gamma}\left(-\frac{i\omega-z}{2\gamma}+i\,\text{sign}(\omega)\sqrt{1-\left(\frac{i\omega-z}{2\gamma}\right)^{2}}\right)~,
\end{equation}
and $S^2(\omega,z)=S^1(\omega,-z)$. As expected, at $z=0$ they reduce to the correlators studied in the previous sections. When $\gamma \to 0$, they reduce to the correlators of a massless fermion in the presence of the $SU(2)$ breaking deformation. From \eqref{eq:exactzdef} we can read off the spectral density, as in \eqref{eq:specdens}:
\begin{equation}\label{eq:zspecdens}
  \rho^1(E,z)=\begin{cases}\frac{1}{\gamma}\sqrt{1-\left(\frac{E-z}{2\gamma}\right)^2}~,& z-2\gamma<E<z+2\gamma~,\\
  0~, &\text{otherwise}
\end{cases}
\end{equation}
with $\rho^2(E,z)=\rho^1(E,-z)$.

\begin{figure}
\begin{center}
\includegraphics[width=0.3\textwidth]{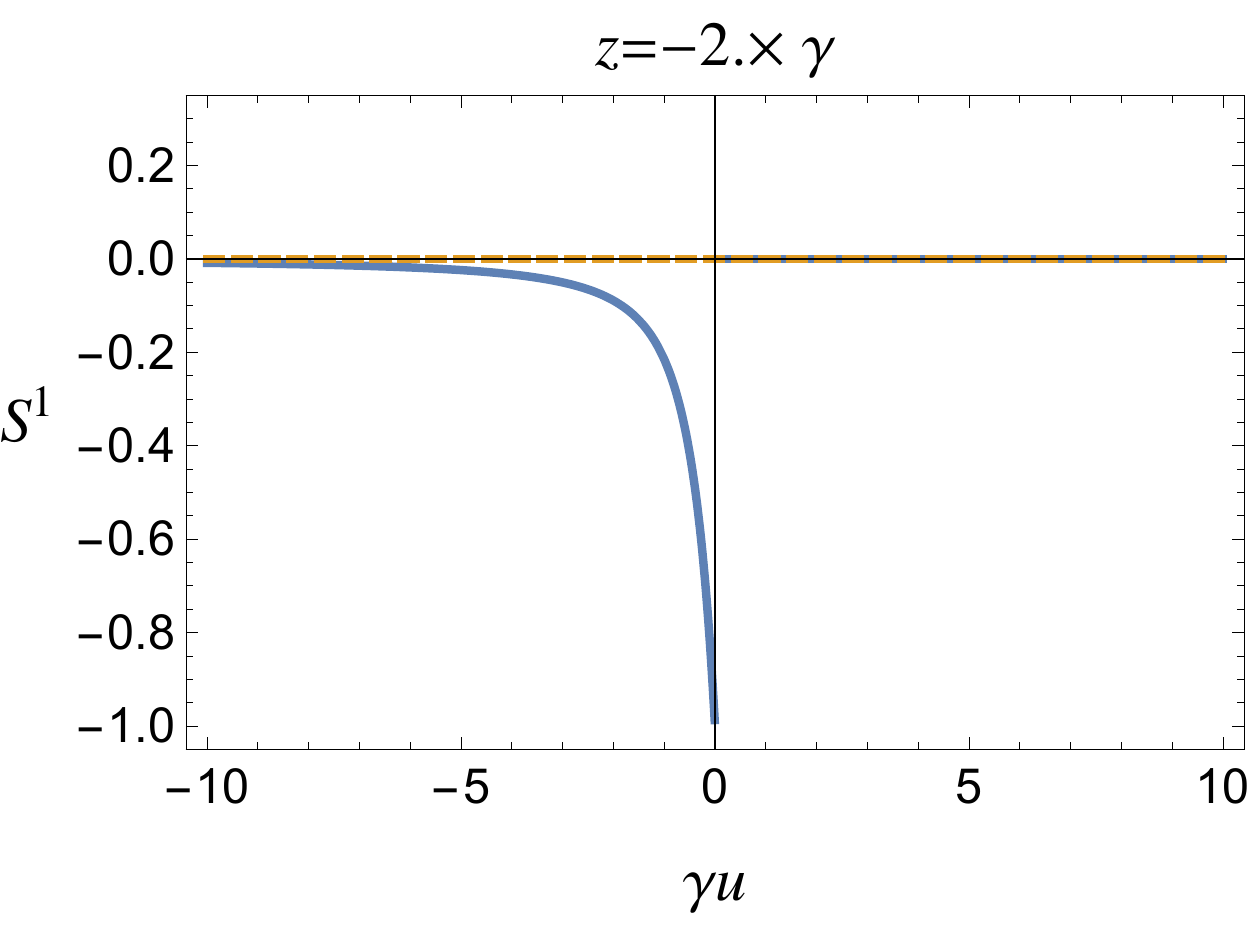}
\includegraphics[width=0.3\textwidth]{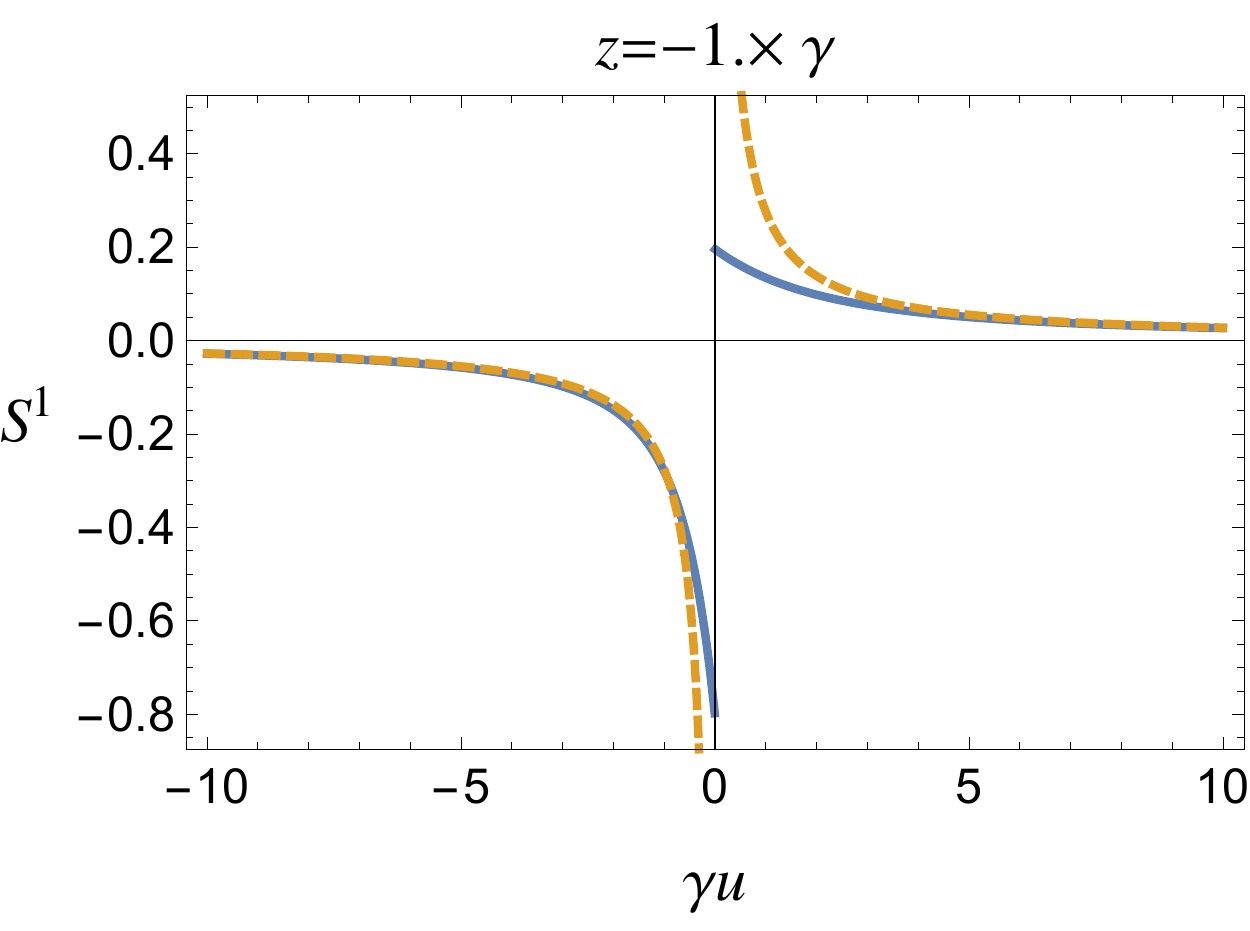}
\includegraphics[width=0.3\textwidth]{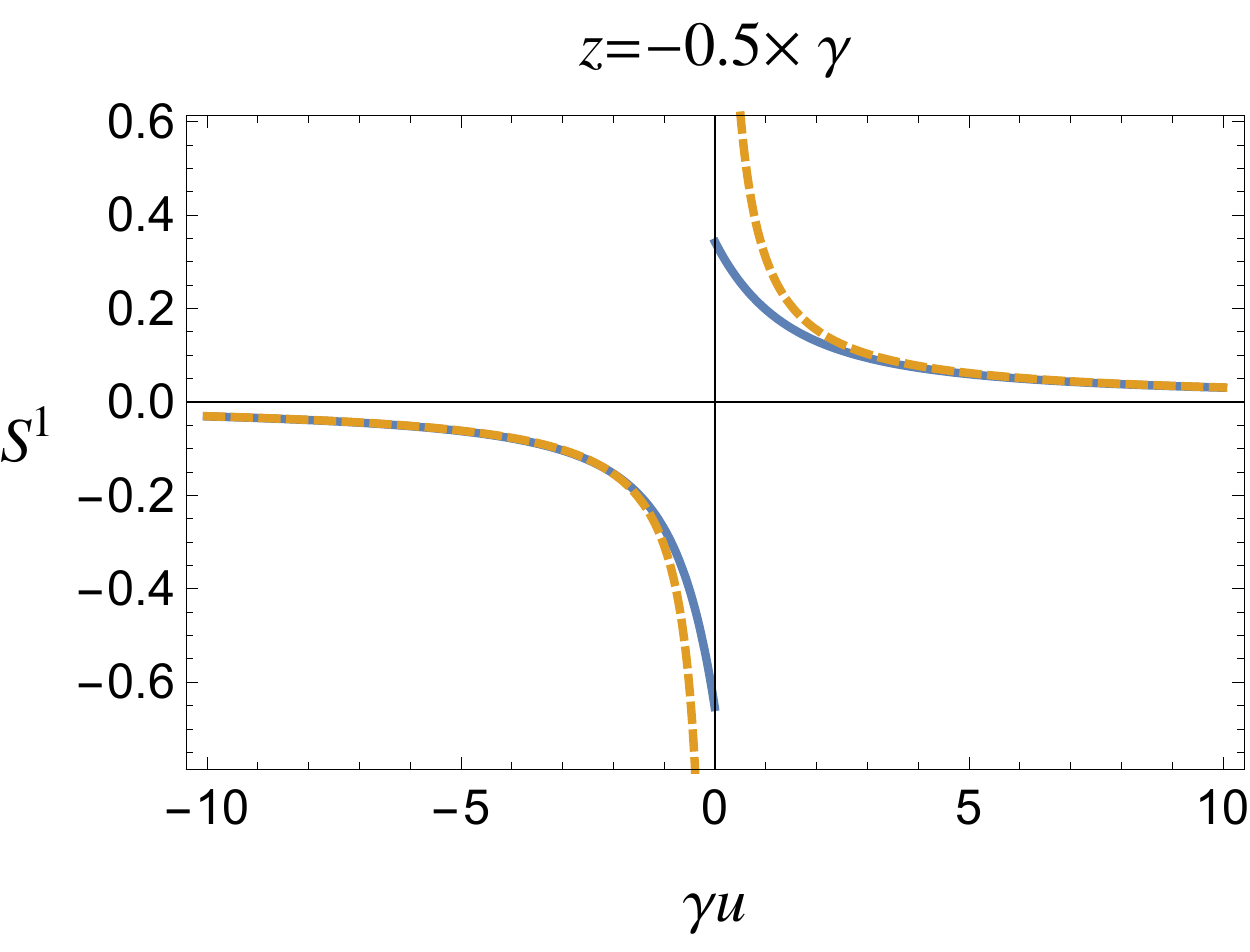}
\includegraphics[width=0.3\textwidth]{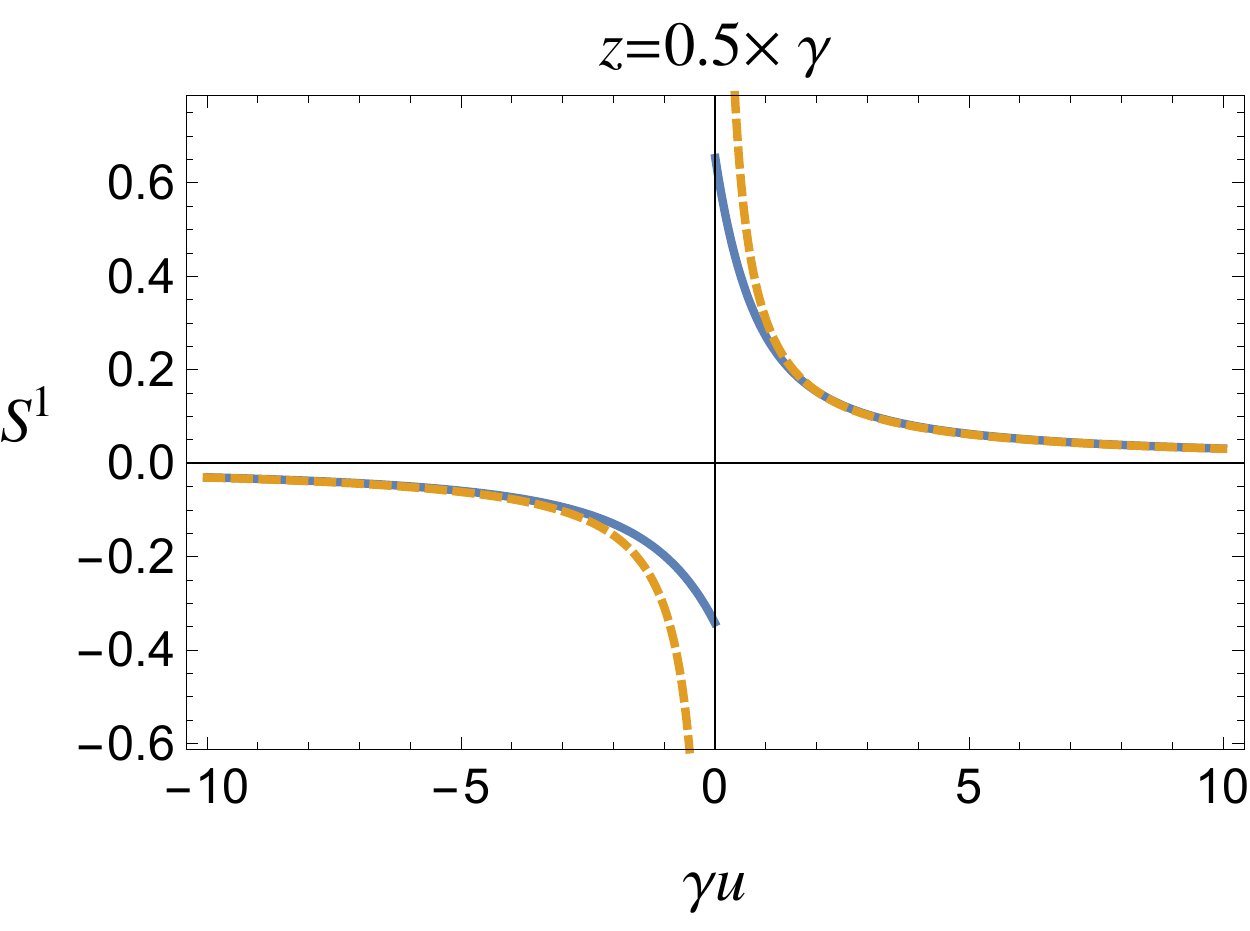}
\includegraphics[width=0.3\textwidth]{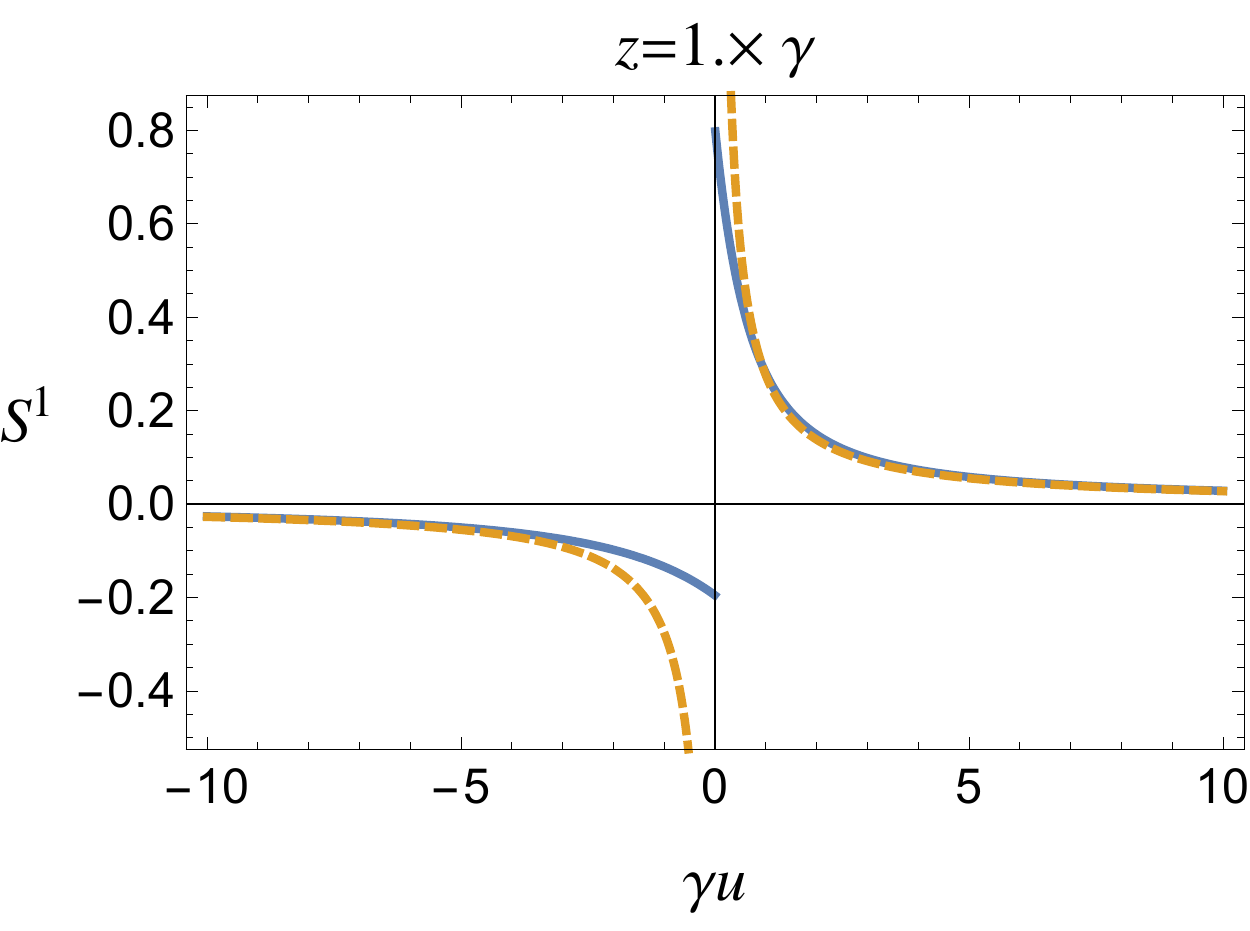}
\includegraphics[width=0.3\textwidth]{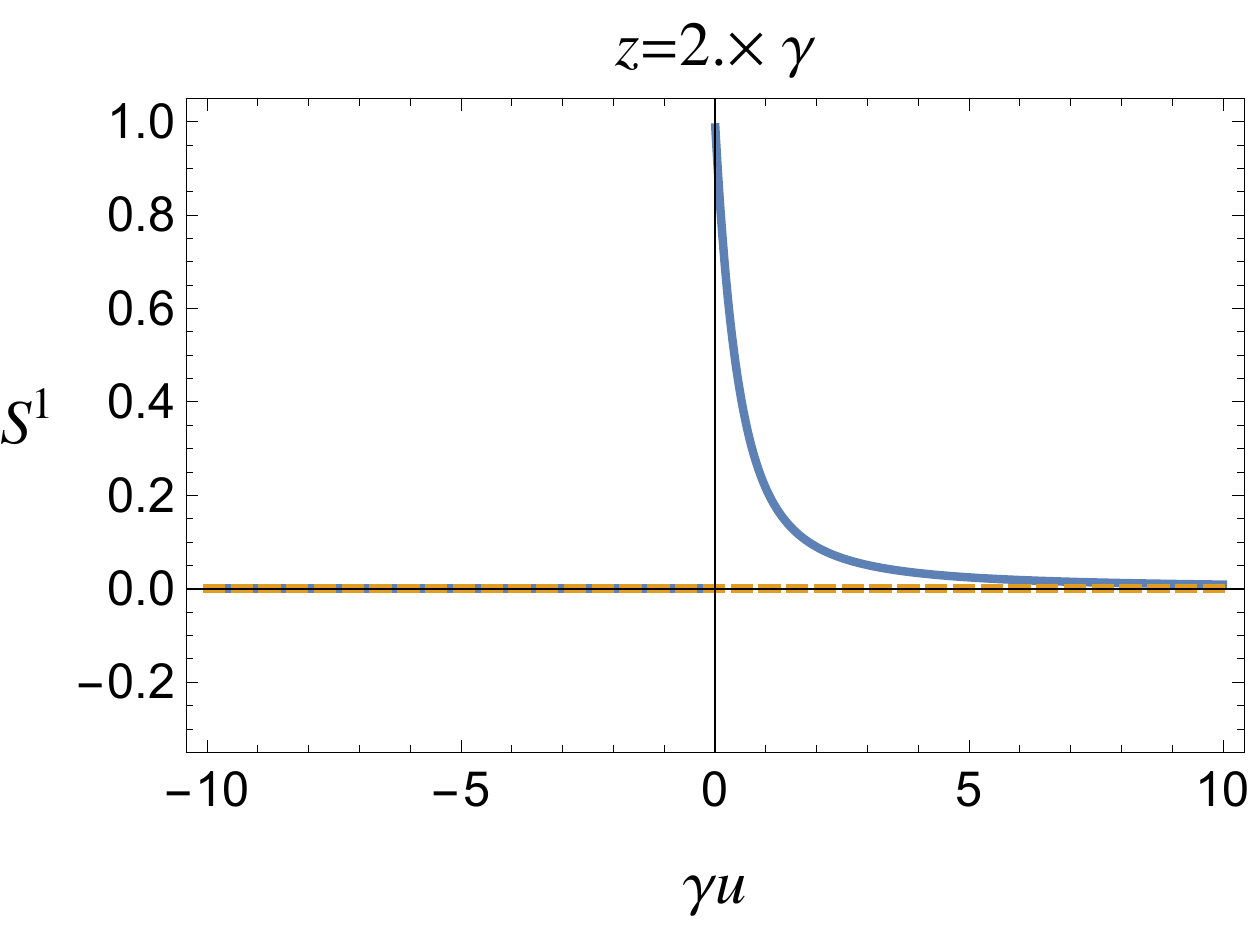}
\caption{The solid blue curves are the numerical Fourier transforms of~(\ref{eq:exactzdef}), $S^1(u)$, $u\equiv \tau-\tau^\prime$, while the dashed orange curves represent the large time separation approximations $S_{\rm low}^1(u)$ in~(\ref{zcorrs}). For $|z|<2\gamma$, $S_{\rm low}^1(u)$ matches the late time behavior of the numerical Fourier transform.}\label{fig:numft}
\end{center}
\end{figure}

At low energy, $\omega\ll \gamma$, and for $|z|< 2\gamma$ the correlators are, at leading order:
\begin{align}
  S_{\text{low}}^1(\omega,z)  =\frac{1}{\gamma}\left(\frac{z}{2\gamma}+i\,\text{sign}(\omega)\sqrt{1-\left(\frac{z}{2\gamma}\right)^2}\right)~,\quad\quad\quad
  S_{\text{low}}^2(\omega,z)  =S_{\text{low}}^1(\omega,-z)~,
\end{align}
which, upon Fourier transforming to Euclidean time, become
\begin{equation}\label{zcorrs}
S_{\text{low}}^1(\tau,\tau') = \frac{1}{\gamma}\left(\delta(\tau-\tau')\, \frac{z}{2\gamma} + \frac{\sqrt{1-\left(\frac{z}{2\gamma}\right)^2}}{ \pi (\tau-\tau')}\right)~, \quad\quad S_{\text{low}}^2(\tau,\tau';z) = S_{\text{low}}^1(\tau,\tau';-z)~.
\end{equation}
Hence for $|z|<2\gamma$ the fermionic correlation functions exhibit the same conformal late time decay as in the undeformed theory, with the same fermionic weight $\Delta_\psi=1/2$. For $z$ within the range $0<|z|<2\gamma$ this deformation is marginal in the IR. We can verify this late time behavior by numerically Fourier transforming the exact solutions \eqref{eq:exactzdef} for any $z$ in the range $|z|<2\gamma$ and compare these to the large time separation approximation. This is shown in figure \ref{fig:numft}. The deformed theory also exhibits a low energy emergent local $U(1)$ symmetry, to be contrasted with the local $SU(2)$ symmetry in the undeformed case. The low energy effective action governing the breaking of these approximate symmetries can be derived as in section \ref{sec:lowenergy}.

Let us now see what happens as we tune $z$ through $z=\pm2\gamma$. The square root has a branch cut that extends along the real $z$ axis for $|z|>2\gamma$. However, at small $|\omega|$ the $i\omega$ inside the square root in (\ref{eq:exactzdef}) acts as an $i\epsilon$ prescription, which crucially depends on the sign of $\omega$, for correctly analytically continuing through $|z|= 2\gamma$. To leading order at low frequencies, we find the solution are real and \emph{constant}:
\begin{equation}
  S_{\text{low}}^1(\omega,z)  =\frac{z}{2\gamma^2}\left(1-\sqrt{1-\left(\frac{2\gamma}{z}\right)^2}\right)~,\quad S_{\text{low}}^2(\omega,z)  =-\frac{z}{2\gamma^2}\left(1-\sqrt{1-\left(\frac{2\gamma}{z}\right)^2}\right)~.
\end{equation}
Thus the low frequency correlations become purely local in Euclidean time $\propto \delta(\tau-\tau')$ for $|z|>2\gamma$, indicating that the fermions barely interact. The correlation function being purely local suggests that the fermions are gapped rather than scale invariant. Indeed we see from the spectral density \eqref{eq:zspecdens} that for $|z|>2\gamma$ there are no states within an open set of $E=0$. In fact, using the spectral representation of the correlator, we can readily compute the Euclidean time expression analytically for $|z|\ge 2\gamma$
\begin{equation}\label{massivecorr}
  S^1(u,z)=\Theta(uz)\frac{e^{-uz}I_1(2|u|\gamma)}{u\gamma}~.
\end{equation}
From the asymptotic form of the Bessel function, we see that this correlator exhibits an exponential decay, e.g. for $z>0$:
\begin{equation} \label{gappedcorr}
  S^1(u\gg0)\sim \frac{e^{-u(z-2\gamma)}}{(u\gamma)^{3/2}}~.
\end{equation}
This late time behavior is similar to the Euclidean time correlation function for a free fermion with mass $\mu$: $G= \Theta(u)e^{-\mu u}$, thus \eqref{massivecorr} allows us to identify the mass gap of this phase as $\mu_{\rm gap}=|z|-2\gamma$. We also note the subleading behavior of the correlator \eqref{gappedcorr} $(u\gamma)^{-3/2}$. This is characteristic of a density of states scaling as $\sqrt{E}$ at low energies, which is precisely the case for $\rho^1(E,z)$ in \eqref{eq:zspecdens} when $z=2\gamma$.\footnote{It is perhaps worth noting that the appearance of a $3/2$ scaling at late times has appeared in several different contexts including \cite{Cotler:2016fpe,Chen:2017yze,Anninos:2016klf}~.}

In summary, at zero temperature turning on $z$ spoils the supersymmetric relation $i \omega Q(\omega) = S^1(\omega) = S^2(\omega)$, as is to be expected. However, finite but small $z$ does not spoil the scaling invariant form of the correlators nor does it shift the conformal weights of the scalars and fermions, which are related by supersymmetry. In this sense, the $SU(2)$ breaking deformation is non-supersymmetric and marginal. When $|z|$ becomes sufficiently large ($>2 \gamma$) the model exhibits a transition to a gapped phase and conformal invariance is lost.

We now proceed to study the thermodynamics and quantum phase structure of the model in the presence of the deformation.

\section{Low temperature phase structure}\label{sec:breakingtherm}

To make contact with the physics of near-extremal rotating black holes, we study the low temperature thermodynamics of the $z$-deformed theory. 
Using the effective action for the fermion two-point function $S$, we find the following finite temperature expression for the angular momentum
\begin{equation}\label{divz}
\frac{1}{Nn} \partial_z \log Z\equiv \beta\langle \hat{J}_z \rangle = \sum_{n\in\mathbb{Z}} \left[ S^1(\omega_n,z) - S^2(\omega_n,z) \right]~.
\end{equation}

\subsection{Small $z$ gapless phase}

Let us start in the regime $|z|<2\gamma$. We can evaluate the sum (\ref{divz}) in the low temperature limit, taking $\omega_n \ll |z|,\gamma$, with $z$ and $\gamma$ large, but of the same order to ensure that $|z|<2\gamma$. The part of the sum related to the low frequency limit at leading order is
\begin{equation}\label{lowTsum}
\beta\langle \hat{J}_z\rangle_{\rm low}= \frac{z}{\gamma^2}\sum_{n \in \mathbb{Z}} \left(1- \frac{ \left| \omega_n \right| }{2\gamma\sqrt{1-\left(\frac{z}{2\gamma}\right)^2}}\right)~.
\end{equation}
 As it stands, the above sum exhibits the same fictitious divergences as~(\ref{Zzero}) and needs to be regularized. We can regulate it using the $\zeta$-function approach adopted in (\ref{zetareg}). With this choice the $\omega_n$-independent term in (\ref{lowTsum}) is regularized to zero. This term is responsible for the  local piece of the correlator proportional to $\delta(\tau-\tau')$. That it does not contribute to the low temperature partition function is an encouraging sign that our regularization is sensible. The $\zeta$-regularized sums give the following low temperature angular momentum
\begin{equation}\label{zetaJ}
 \beta \langle \hat{J}_z \rangle_{\rm low}  = - \frac{\pi z}{12\beta\gamma^3\sqrt{1-\left(\frac{z}{2\gamma}\right)^2}}~,\end{equation}
where we have reintroduced explicit factors of $\beta$. We note a transition at $z=\pm2\gamma$, where the angular momentum diverges. That it diverges is an artifact of the large $N$ limit, given that at any finite $N$ there is no state with infinite angular momentum. At finite $N$ the above transition is smoothed out.

\begin{figure}
\begin{center}
{\includegraphics[width=0.4\textwidth]{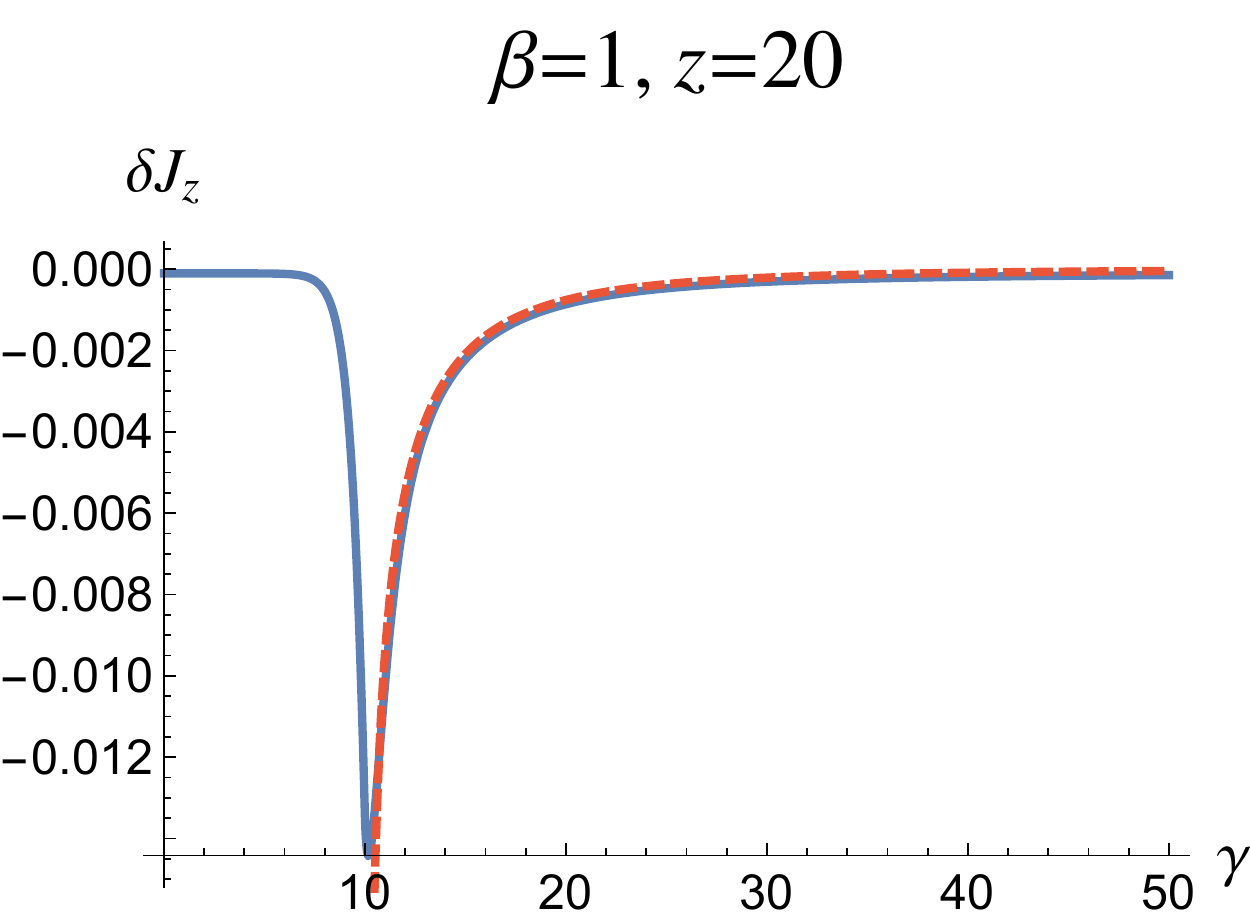}
\quad\quad\includegraphics[width=0.45\textwidth]{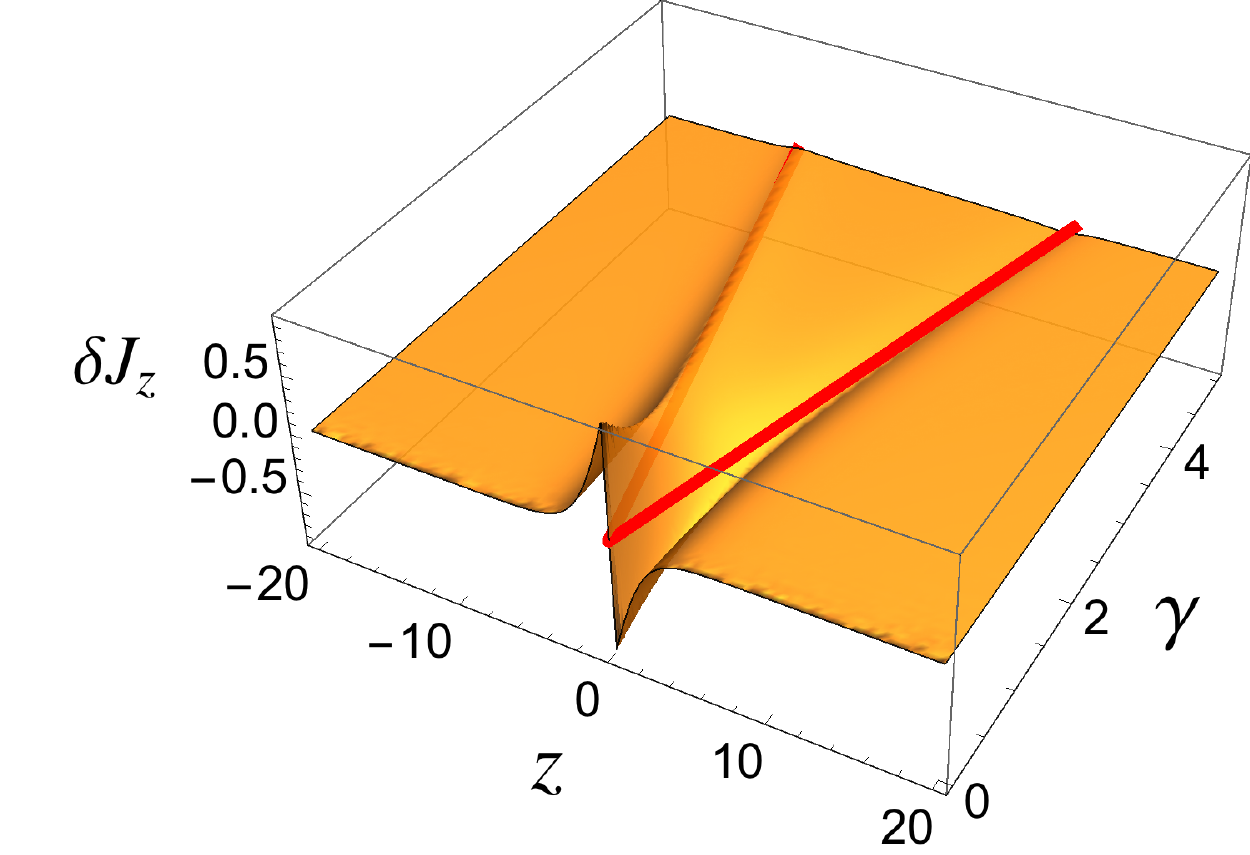}}
\end{center}
\caption{\emph{Left}: Comparison between $\delta J_z \equiv \left( \langle \hat{J}_z \rangle_\beta -  \langle \hat{J}_z \rangle_0 \right)$ (solid-blue) and $\langle \hat{J}_z\rangle_{\rm low}$ (dashed-red) given in (\ref{zetaJ}), for $z=20$ (in units where $\beta=1$), with a cutoff on the sum given by $n_c=\pm2\times10^4$. Note the sharp transition at $\gamma=z/2$. \emph{Right}: $\delta J_z$ vs. $z$ and $\gamma$. The red line indicates the locus $|z|=2\gamma$.}\label{trans}
\end{figure}

We can integrate (\ref{zetaJ}) with respect to $z$ to obtain the $z$-corrected low temperature free energy
\begin{equation}
\frac{1}{Nn}\log Z[\beta,z] =-\beta \left(E_0(z)-z\langle\hat{J}_z\rangle_0\right)+ \frac{\pi}{3\beta\gamma}\sqrt{1-\left(\frac{z}{2\gamma}\right)^2}~.
\end{equation}
What remains is to compute the contributions at $T=0$ to energy and angular momentum. At exactly zero temperature, we can perform the continuous $\omega$-integral
\begin{equation}
\langle\hat{J}_z\rangle_0 = \int_{\mathbb{R}} \frac{d\omega}{2\pi} \left( S^1(\omega,z)-S^2(\omega,z) \right) ~.
\end{equation}
which computes the angular momentum (divided by $Nn$). Remarkably, this integral can be performed exactly, for the full $S^a(\omega, z)$ solution, yielding the following result for $|z|< 2\gamma$:
\begin{equation}\label{zeroJ}
\langle\hat{J}_z\rangle_0 =\frac{2}{\pi}\left(\sin^{-1}\left(\frac{z}{2\gamma}\right)+\frac{z}{2\gamma}\sqrt{1-\left(\frac{z}{2\gamma}\right)^2}\right)
\end{equation}
The minus sign in the small temperature contribution (\ref{zetaJ}) means that the angular momentum at small temperatures is slightly less than that at vanishing temperatures. As a check on (\ref{zetaJ}), we compare it to a numerical evaluation of the full sum (\ref{lowTsum}) with (\ref{zeroJ}) subtracted. They are in good agreement, as shown in figure \ref{trans}.

Notice that in the limit $z\to \pm2\gamma$, the zero temperature angular momentum $\langle \hat{J}_z \rangle_0$ tends to $\pm 1$. At $z=\pm 2\gamma$, the total zero temperature angular momentum becomes $\pm N$, i.e. all spins pointing up or down. A simple computation for the ground state energy, analogous to the $z=0$ case, yields
\begin{equation}
 E_0(z) = -\frac{8\gamma}{3 \pi }\left(1-\left(\frac{z}{2\gamma}\right)^2\right)^{3/2}~.
\end{equation}

\subsection{Large $z$ gapped phase}

When $|z|>2\gamma$ the fermionic correlation functions become purely local in the low frequency limit. Thus, the low frequency contribution to the partition function no longer has a non-trivial frequency dependence, it is purely constant. Using our $\zeta$-function regulator, we simply evaluate
\begin{equation}
\partial_z \log Z_{\text{low}}[\beta,z] = 0~.
\end{equation}
Indeed, for $|z|>2\gamma$, the low temperature thermodynamics is insensitive to $z$. As we shall see, this reflects the fact that the fermions become gapped at low temperatures. The integral at exactly zero temperature can also be performed, yielding
\begin{equation}
\langle \hat{J}_z \rangle_0 = \text{sign}(z)~.
\end{equation}
The angular momentum for $|z|>2\gamma$ is constant, and only depends on the sign of $z$. Multiplying by $N$ gives us the total angular momentum, meaning that for all $|z|>2\gamma$ the total angular momentum is $\pm N$ in the $\hat{z}$-direction, so all spins are pointing up or down.

The low temperature partition function for $|z| > 2\gamma$, including the zero temperature piece, is given to leading order by:
\begin{equation}
\frac{1}{Nn} \, \log Z[\beta,z] = \beta |z| + \ldots
\end{equation}
Notice that for $|z|>2\gamma$, both the entropy and the energy are vanishing to leading order in the low temperature expansion.

\subsection{Phase transition}

\begin{figure}
\begin{center}
{\includegraphics[width=0.5\textwidth]{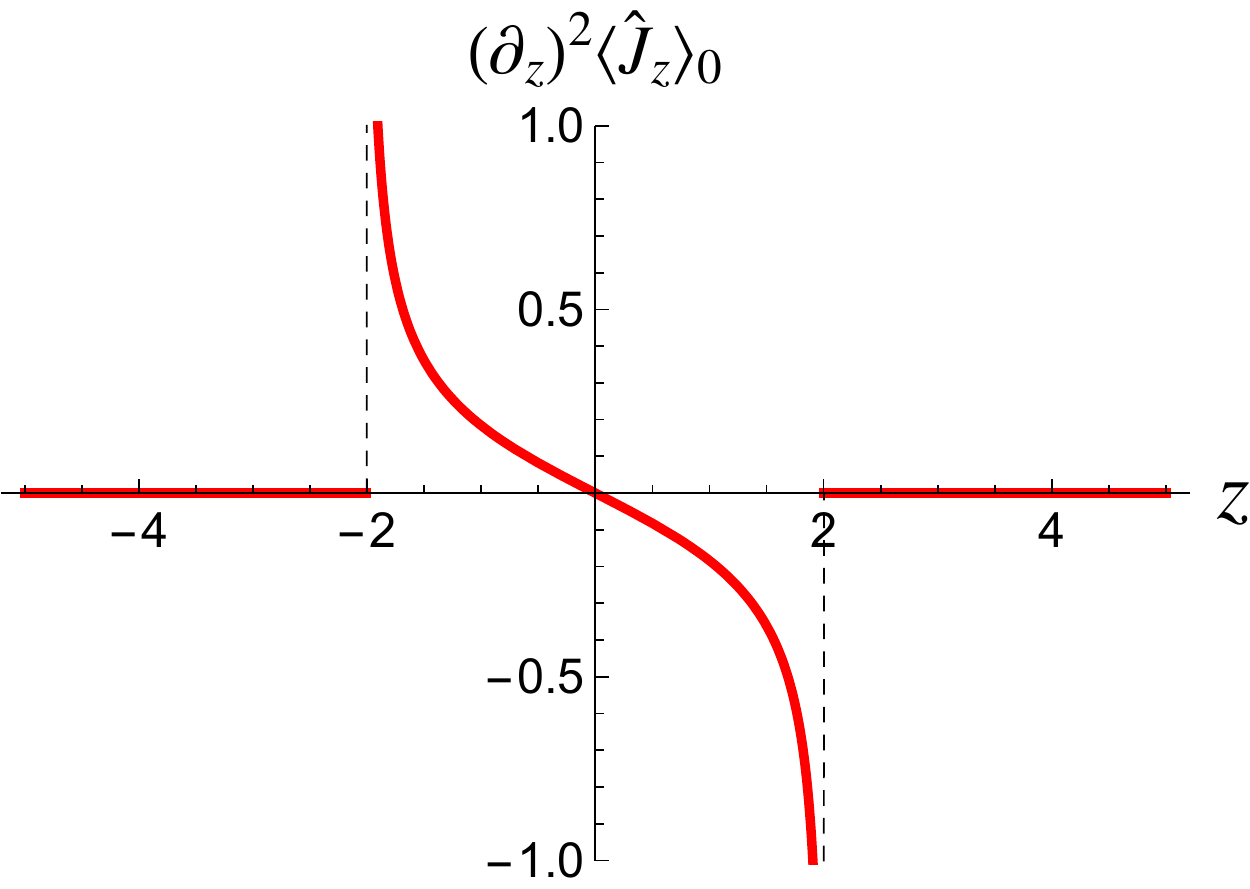}}
\end{center}
\caption{$d^2\langle \hat{J}_z \rangle_0/dz^2$ vs. $z$, for $\gamma=1$.}\label{PT}
\end{figure}

At exactly zero temperature, the transition between $|z|<2\gamma$ and $|z|>2\gamma$ is not smooth. In particular, coming from $|z|<2\gamma$ we can compute
\begin{equation}\label{divz2}
\frac{d^2}{dz^2} \langle \hat{J}_z \rangle_0  = -\frac{z}{2\pi\gamma^3 \sqrt{1-\left(\frac{z}{2\gamma}\right)^2}}~.
\end{equation}
On the other hand, coming from $|z|>2\gamma$ we obtain a vanishing second derivative. Both $\langle J_z \rangle_0$ and its first derivative are continuous as a function of $z$ across the transition, indicating that we have a higher order quantum phase transition. We plot this in figure \ref{PT}. The critical exponent of $\partial_z^2 \langle \hat{J}_z \rangle_0$ is $-1/2$ can be obtained from the behavior of (\ref{divz2}) near $|z_c| = 2\gamma$.

At small but non-vanishing temperatures there is another discontinuity that we observe. For $|z|<2\gamma$ we have:
\begin{equation}\label{divz1}
\beta\langle \hat{J}_z  \rangle_{\text{low}}  = -\frac{\pi z}{12\beta \gamma^3 \sqrt{1-\left(\frac{z}{2\gamma}\right)^2}}~.
\end{equation}
whereas for $|z|>2\gamma$, we find $\langle \hat{J}_z  \rangle_{\text{low}}  = 0$. Interestingly, the strength of the discontinuity at exactly zero temperature (\ref{divz2}) is subleading compared to that from the sub-leading piece in the small temperature expansion (\ref{divz1}), which is a divergence of $\langle \hat{J}_z \rangle_\beta$ itself, with critical exponent $-1/2$.

We emphasize here the different orders of limits that we have chosen. In both cases, we are first taking the large $N$ limit. But then we must decide whether we are taking the zero temperature limit before or after the $z\to\pm 2\gamma$ limit. Depending on this order of limits, the phase structure is either like the zero temperature one or the low temperature one.
\newline\newline\newline
To summarize, for small values of $z$, the system is disordered and gapless, and the dominant low temperature states have small angular momentum. As $z$ increases, we are biasing an increasing number of states to have angular momentum in the $\hat{z}$-direction. Finally, at $|z|=2\gamma$, all the spins are pointing in the same direction and the system enters an ordered, gapped phase. The presence of a non-analyticity in the system is an artifact of taking the large $N$ limit first. At finite but large $N$ the transitions described above are smooth.

\subsection{Comparison to rotating black holes}

Our quantum system exhibits a coupling $z$ that breaks $SU(2)$, while preserving $SL(2,\mathbb{R})$ up to some finite value. Moreover, the partition function exhibits a non-analytic behavior in $z$ at the transition point. These are common features between the quantum mechanics and the black hole system, where we can identify $z$ with the angular velocity of the horizon $\Omega$ (that acts as a chemical potential for the angular momentum). These are encouraging signs that the black hole picture may have a purely quantum mechanical interpretation. We find it curious, and perhaps interesting, that for the particular choice of clock discussed at the end of section \ref{BHPT}, the non-analytic behavior $C_{BH} \sim \sqrt{1-8Q^2\Omega^2}$ of the black hole specific heat is precisely that of of the deformed quantum system $C_{QM} \sim \sqrt{1-(z/2\gamma)^2}$. However, the non-analyticities in the black hole case depend on how the near-horizon clock is defined, and we leave it to future work to sharpen this analogy.
We also notice that the leading contribution in the limit $z\to 2\gamma$ to the free energy at exactly zero temperature displays the same non-analytic behavior as the free energy of the black hole at $T=0$, once we define an appropriate chemical potential for the angular momentum (see appendix \ref{appLeft}). However the derivatives of the free energy differ in the two cases.

\section{Outlook on future directions}\label{sec:discussion}

The main quantitative results of the paper and their relation to black hole physics were presented at the end of the previous section. Here we conclude with an outlook on future directions.

\subsection{Five dimensional black holes}
Five-dimensional black holes differ in several interesting ways from their four-dimensional counterparts. For instance, there is no analog of the no hair theorem in five-dimensional asymptotically flat space. The most symmetric solution is non-rotating and preserves the $SO(4)\cong SU(2)_L\times SU(2)_R$ spatial symmetry group. When extremal, the solutions of five-dimensional Einstein-Maxwell theory have also been shown to generally exhibit an $SL(2,\mathbb{R})$ symmetry \cite{Kunduri:2013ana}. There is a large class of such $SL(2,\mathbb{R})$ invariant near-horizon geometries for these black holes which can be classified by the symmetries they preserve. Depending on the charges and angular momenta the preserved symmetries are either: $SO(4)$, $SU(2)\times U(1)$ or a $U(1)\times U(1)$ rotational subgroup. It would be interesting to interpret this in the language of the  quantum mechanical models that we have considered. Since $SO(4)\cong SU(2)_L\times SU(2)_R$, we can start with two separate sets of fermions $\psi^a_{\mathcal{L}}$ and $\psi^a_{\mathcal{R}}$ transforming under $SU(2)_{\mathcal{L}} \times SU(2)_{\mathcal{R}}$. Then it is easy to implement any breaking pattern by turning on chemical potentials for different fermion bilinears. In the gravity theory, an interesting possibility is to turn on two angular momenta of the same magnitude $|\vec J_L|=|\vec J_R|$. Then it can be shown that the residual symmetry of the solution is $SU(2)$~\cite{Kunduri:2013ana}. This suggests that this configuration corresponds to fermions excited purely in the $SU(2)_{\mathcal{L}}$ (or  $SU(2)_{\mathcal{R}}$) sector (see \cite{Maldacena:1997ih} for a related discussion).

\subsection{Interacting models and ground state degeneracy}

An important requirement of more realistic models is a large ground state degeneracy. Our models do not exhibit this crucial feature for the phenomenology of extremal black holes. One way to achieve a large degeneracy of ground states would be to consider a cubic rather than quadratic superpotential. This was precisely the case considered in \cite{Anninos:2016szt}, where it was argued that the supersymmetric saddle has an IR limit where the fermionic scaling dimension is $\Delta_\psi  = 1/2$. A cubic superpotential is already enough to have a (supersymmetric) ground state degeneracy that grows exponentially in $N$ \cite{Denef:2007vg}. Moreover, this theory gives rise to more intricate higher point functions then the model under consideration. We hope to report on this case in the future. One could also consider SYK like systems, built  purely out of fermions, with an additional $SU(2)$ index.

\subsection{Superradiance}

Rotating black holes in asymptotically flat space are known to superradiate. Classically, when we send an incident wave toward the horizon it may scatter back out with more energy. Quantum mechanically, this can occur spontaneously and is a rotational analog of Hawking radiation. Any quantum mechanical dual of a rotating black hole must somehow exhibit this feature \cite{Bredberg:2009pv}. Superradiance connects the near-horizon geometry to the asymptotic region, since it induces non-vanishing flux at the boundary of the near-horizon geometry. From the point of view of AdS$_2$, superradiance of an extremally rotating black hole manifests itself in terms of Schwinger-pair production of highly charged particles in a background $\bold{E}$-field (see for example \cite{Anninos:2009jt}). These particles cannot be contained in the AdS$_2$ near-horizon region and leak out of its boundary \cite{Pioline:2005pf}. This manifests itself in terms of complex scaling dimensions of the form $\Delta = 1/2 + i \nu$, with $\nu$ real. Interestingly, $SL(2,\mathbb{R})$ allows for unitary irreducible representations that have such complex scaling dimensions. It would be interesting to construct $SL(2,\mathbb{R})$ invariant models with operators transforming under these irreducible representations and understand if they are related to superradiant modes.

\subsection{Rotating de Sitter horizons}

There is an interesting limit where a rotating black hole in a de Sitter space has the size of the cosmological horizon. In this case, one finds a $SL(2,\mathbb{R})\times U(1)$ invariant solution to the Einstein equations endowed with a positive cosmological constant \cite{Booth:1998gf,Anninos:2009yc,Anninos:2010gh}. When the rotation is turned off it reduces to the dS$_2\times S^2$ Nariai geometry, and the $U(1)$ is enhanced to an $SO(3)$. These features are remarkably similar to those of the Kerr-Newman case, although in de Sitter there is no Maxwell field necessary. It would be interesting to understand whether our discussion is also relevant for these geometries.

\section*{Acknowledgements}

It is a great pleasure to thank Frederik Denef, Diego Hofman, and Benson Way for useful discussions. R.T.D. would like to thank Masha Baryakhtar for a useful discussion about X-ray measurements of astrophysical black holes. D.A. is supported by the AMIAS and NSF. T.A. is supported by the Natural Sciences and Engineering Research Council of Canada, and by grant 376206 from the Simons Foundation. R.T.D. is supported by Swiss National Science Foundation contract 200020-169696 and the Sinergia network CRSII2-160814.

\appendix

\section{Left moving chemical potential}\label{appLeft}
In this appendix we  consider an ensemble for extremal ($T=0$) rotating black holes different than the one studied in the main text. The thermodynamic potential $\mathcal{G}$ (again at fixed $Q$) of interest is:
\begin{equation}
\mathcal{G} = J- T_L \,  S~,
\end{equation}
 where the parameter $\beta_L\equiv 1/T_L$ pays the role of a chemical potential for $J$. One can show that $T_L$ is the effective temperature felt by left-moving modes near the extremal horizon, upon tracing out the interior region in the Frolov-Thorne vacuum state \cite{Frolov:1989jh}.  Equilibrium is attained when $\delta \mathcal{G} = 0$ with respect to variations of $J$ and $S$, thus defining a zero-temperature first law. The zero temperature partition function $Z$ is defined as
\begin{equation}
\log Z  \equiv -\beta_L \mathcal{G}~ \quad \implies \quad \partial_{\beta_L} \log Z = -J(\beta_L)~.
\end{equation}
At $T=0$ we can parametrize the black hole entropy in terms of $J$ as
\begin{equation}
S = {\pi}\sqrt{4J^2+Q^4}~,
\end{equation}
and the $T=0$ first law gives the equilibrium expression for $T_L$:
\begin{equation}\label{1stlaw}
\left.\frac{\delta S}{\delta J}\right|_{Q \text{ fixed}}  = \beta_L= \frac{4 \pi  J}{\sqrt{4 J^2+Q^4}}~.
\end{equation}
 At $Q = 0$, this reduces to $T_L = \pm 1/2\pi$. We can relate $\beta_L$ to the angular velocity at the horizon
  \begin{equation}
   \beta_L^{\pm}=\pi\,\text{sign}(\Omega)\sqrt{(1\mp s)(3\pm s)}~,
 \end{equation}
 with $s$ defined in \eqref{eq:sdef}. Along the the positive branch, $\beta_L^+ = 0$ at $\Omega = 0$ whereas $\beta_L^- = \pm 2\pi$ at $\Omega = 0$ in the negative branch. The branches meet at at $\Omega = \pm 1/\sqrt{8}Q$ where $\beta_L = \pm\sqrt{3}\pi$ .

We can invert \eqref{1stlaw} and obtain $J(\beta_L)$:
\begin{equation}
J(\beta_L) =   \frac{Q^2 \beta_L}{2} \, \frac{1}{\sqrt{4\pi^2- \beta_L^2}}~.
\end{equation}
So the thermodynamic potential is given as function of $T_L$:
\begin{equation}
\mathcal{G} =  - \frac{Q^2}{2\beta_L} \sqrt{ 4 \pi^2-\beta_L^2}~,
\end{equation}
or equivalently the partition function is given by:
\begin{equation}
 \log Z =  \frac{Q^2}{2} \sqrt{ 4 \pi^2-\beta_L^2}~.
\end{equation}
Notice the non-analyticity at $\beta_L = \pm 2\pi$. At these values of $\beta_L$ we have that $J(\beta_L)$ diverges and the divergence has a (quantum) critical exponent of one-half. We would like to interpret this as a quantum phase transition. It is different from the $Q^2\Omega^2 = 1/8$ divergence we found previously, where the angular momentum did not diverge.

\bibliographystyle{utphys}
\bibliography{rotatingrefs}{}

\providecommand{\href}[2]{#2}\begingroup\raggedright\begin{thebibliography}{10}

\bibitem{Reynolds:2013qqa}
C.~S. Reynolds, ``{Measuring Black Hole Spin using X-ray Reflection
  Spectroscopy},'' \href{http://dx.doi.org/10.1007/s11214-013-0006-6}{{\em
  Space Sci. Rev.} {\bfseries 183} no.~1-4, (2014) 277--294},
\href{http://arxiv.org/abs/1302.3260}{{\ttfamily arXiv:1302.3260
  [astro-ph.HE]}}.

\bibitem{McClintock:2013vwa}
J.~E. McClintock, R.~Narayan, and J.~F. Steiner, ``{Black Hole Spin via
  Continuum Fitting and the Role of Spin in Powering Transient Jets},''
  \href{http://dx.doi.org/10.1007/s11214-013-0003-9}{{\em Space Sci. Rev.}
  {\bfseries 183} (2014) 295--322},
\href{http://arxiv.org/abs/1303.1583}{{\ttfamily arXiv:1303.1583
  [astro-ph.HE]}}.

\bibitem{Reynolds:2013rva}
C.~S. Reynolds, ``{The Spin of Supermassive Black Holes},''
  \href{http://dx.doi.org/10.1088/0264-9381/30/24/244004}{{\em Class. Quant.
  Grav.} {\bfseries 30} (2013) 244004},
\href{http://arxiv.org/abs/1307.3246}{{\ttfamily arXiv:1307.3246
  [astro-ph.HE]}}.

\bibitem{Strominger:1998yg}
A.~Strominger, ``{AdS(2) quantum gravity and string theory},''
  \href{http://dx.doi.org/10.1088/1126-6708/1999/01/007}{{\em JHEP} {\bfseries
  01} (1999) 007},
\href{http://arxiv.org/abs/hep-th/9809027}{{\ttfamily arXiv:hep-th/9809027
  [hep-th]}}.

\bibitem{Sen:2011cn}
A.~Sen, ``{State Operator Correspondence and Entanglement in $AdS_2/CFT_1$},''
  \href{http://dx.doi.org/10.3390/e13071305}{{\em Entropy} {\bfseries 13}
  (2011) 1305--1323},
\href{http://arxiv.org/abs/1101.4254}{{\ttfamily arXiv:1101.4254 [hep-th]}}.

\bibitem{Bardeen:1999px}
J.~M. Bardeen and G.~T. Horowitz, ``{The Extreme Kerr throat geometry: A Vacuum
  analog of AdS(2) x S**2},''
  \href{http://dx.doi.org/10.1103/PhysRevD.60.104030}{{\em Phys. Rev.}
  {\bfseries D60} (1999) 104030},
\href{http://arxiv.org/abs/hep-th/9905099}{{\ttfamily arXiv:hep-th/9905099
  [hep-th]}}.

\bibitem{Bertotti:1959pf}
B.~Bertotti, ``{Uniform electromagnetic field in the theory of general
  relativity},''
\href{http://dx.doi.org/10.1103/PhysRev.116.1331}{{\em Phys. Rev.} {\bfseries
  116} (1959) 1331}.

\bibitem{Robinson:1959ev}
I.~Robinson, ``{A Solution of the Maxwell-Einstein Equations},''
{\em Bull. Acad. Pol. Sci. Ser. Sci. Math. Astron. Phys.} {\bfseries 7} (1959)
  351--352.

\bibitem{Guica:2008mu}
M.~Guica, T.~Hartman, W.~Song, and A.~Strominger, ``{The Kerr/CFT
  Correspondence},'' \href{http://dx.doi.org/10.1103/PhysRevD.80.124008}{{\em
  Phys. Rev.} {\bfseries D80} (2009) 124008},
\href{http://arxiv.org/abs/0809.4266}{{\ttfamily arXiv:0809.4266 [hep-th]}}.

\bibitem{Anninos:2008fx}
D.~Anninos, W.~Li, M.~Padi, W.~Song, and A.~Strominger, ``{Warped AdS(3) Black
  Holes},'' \href{http://dx.doi.org/10.1088/1126-6708/2009/03/130}{{\em JHEP}
  {\bfseries 03} (2009) 130},
\href{http://arxiv.org/abs/0807.3040}{{\ttfamily arXiv:0807.3040 [hep-th]}}.

\bibitem{Castro:2009jf}
A.~Castro and F.~Larsen, ``{Near Extremal Kerr Entropy from AdS(2) Quantum
  Gravity},'' \href{http://dx.doi.org/10.1088/1126-6708/2009/12/037}{{\em JHEP}
  {\bfseries 12} (2009) 037},
\href{http://arxiv.org/abs/0908.1121}{{\ttfamily arXiv:0908.1121 [hep-th]}}.

\bibitem{Detournay:2012pc}
S.~Detournay, T.~Hartman, and D.~M. Hofman, ``{Warped Conformal Field
  Theory},'' \href{http://dx.doi.org/10.1103/PhysRevD.86.124018}{{\em Phys.
  Rev.} {\bfseries D86} (2012) 124018},
\href{http://arxiv.org/abs/1210.0539}{{\ttfamily arXiv:1210.0539 [hep-th]}}.

\bibitem{Bardeen:1983rv}
W.~A. Bardeen, M.~Moshe, and M.~Bander, ``{Spontaneous Breaking of Scale
  Invariance and the Ultraviolet Fixed Point in O($n$) Symmetric ($\phi^{6}$ in
  Three-Dimensions) Theory},''
\href{http://dx.doi.org/10.1103/PhysRevLett.52.1188}{{\em Phys. Rev. Lett.}
  {\bfseries 52} (1984) 1188}.

\bibitem{kitaev}
A.~Kitaev, ``{A simple model of quantum holography},'' {\em KITP strings
  seminar and Entanglement program} (April 7, 2015 and May 27, 2015) .
  \url{http://online.kitp.ucsb.edu/online/entangled15/}, and
  \url{http://online.kitp.ucsb.edu/online/entangled15/kitaev2/}.

\bibitem{Polchinski:2016xgd}
J.~Polchinski and V.~Rosenhaus, ``{The Spectrum in the Sachdev-Ye-Kitaev
  Model},'' \href{http://dx.doi.org/10.1007/JHEP04(2016)001}{{\em JHEP}
  {\bfseries 04} (2016) 001},
\href{http://arxiv.org/abs/1601.06768}{{\ttfamily arXiv:1601.06768 [hep-th]}}.

\bibitem{Maldacena:2016hyu}
J.~Maldacena and D.~Stanford, ``{Remarks on the Sachdev-Ye-Kitaev model},''
  \href{http://dx.doi.org/10.1103/PhysRevD.94.106002}{{\em Phys. Rev.}
  {\bfseries D94} no.~10, (2016) 106002},
\href{http://arxiv.org/abs/1604.07818}{{\ttfamily arXiv:1604.07818 [hep-th]}}.

\bibitem{Gross:2016kjj}
D.~J. Gross and V.~Rosenhaus, ``{A Generalization of Sachdev-Ye-Kitaev},''
  \href{http://dx.doi.org/10.1007/JHEP02(2017)093}{{\em JHEP} {\bfseries 02}
  (2017) 093},
\href{http://arxiv.org/abs/1610.01569}{{\ttfamily arXiv:1610.01569 [hep-th]}}.

\bibitem{Sachdev:2015efa}
S.~Sachdev, ``{Bekenstein-Hawking Entropy and Strange Metals},''
  \href{http://dx.doi.org/10.1103/PhysRevX.5.041025}{{\em Phys. Rev.}
  {\bfseries X5} no.~4, (2015) 041025},
\href{http://arxiv.org/abs/1506.05111}{{\ttfamily arXiv:1506.05111 [hep-th]}}.

\bibitem{Anninos:2016szt}
D.~Anninos, T.~Anous, and F.~Denef, ``{Disordered Quivers and Cold Horizons},''
  \href{http://dx.doi.org/10.1007/JHEP12(2016)071}{{\em JHEP} {\bfseries 12}
  (2016) 071},
\href{http://arxiv.org/abs/1603.00453}{{\ttfamily arXiv:1603.00453 [hep-th]}}.

\bibitem{Jevicki:2016bwu}
A.~Jevicki, K.~Suzuki, and J.~Yoon, ``{Bi-Local Holography in the SYK Model},''
  \href{http://dx.doi.org/10.1007/JHEP07(2016)007}{{\em JHEP} {\bfseries 07}
  (2016) 007},
\href{http://arxiv.org/abs/1603.06246}{{\ttfamily arXiv:1603.06246 [hep-th]}}.

\bibitem{Anninos:2013nra}
D.~Anninos, T.~Anous, P.~de~Lange, and G.~Konstantinidis, ``{Conformal quivers
  and melting molecules},''
  \href{http://dx.doi.org/10.1007/JHEP03(2015)066}{{\em JHEP} {\bfseries 03}
  (2015) 066},
\href{http://arxiv.org/abs/1310.7929}{{\ttfamily arXiv:1310.7929 [hep-th]}}.

\bibitem{Frolov:1989jh}
V.~P. Frolov and K.~S. Thorne, ``{Renormalized Stress - Energy Tensor Near the
  Horizon of a Slowly Evolving, Rotating Black Hole},''
\href{http://dx.doi.org/10.1103/PhysRevD.39.2125}{{\em Phys. Rev.} {\bfseries
  D39} (1989) 2125--2154}.

\bibitem{Maldacena:2016upp}
J.~Maldacena, D.~Stanford, and Z.~Yang, ``{Conformal symmetry and its breaking
  in two dimensional Nearly Anti-de-Sitter space},''
  \href{http://dx.doi.org/10.1093/ptep/ptw124}{{\em PTEP} {\bfseries 2016}
  no.~12, (2016) 12C104},
\href{http://arxiv.org/abs/1606.01857}{{\ttfamily arXiv:1606.01857 [hep-th]}}.

\bibitem{Denef:2002ru}
F.~Denef, ``{Quantum quivers and Hall / hole halos},''
  \href{http://dx.doi.org/10.1088/1126-6708/2002/10/023}{{\em JHEP} {\bfseries
  10} (2002) 023},
\href{http://arxiv.org/abs/hep-th/0206072}{{\ttfamily arXiv:hep-th/0206072
  [hep-th]}}.

\bibitem{Yoon:2017nig}
J.~Yoon, ``{SYK Models and SYK-like Tensor Models with Global Symmetry},''
\href{http://arxiv.org/abs/1707.01740}{{\ttfamily arXiv:1707.01740 [hep-th]}}.

\bibitem{sachdev2011}
S.~Sachdev, {\em Quantum phase transitions}.
\newblock Cambridge University Press, Cambridge, second ed.~ed., 2011.

\bibitem{Davison:2016ngz}
R.~A. Davison, W.~Fu, A.~Georges, Y.~Gu, K.~Jensen, and S.~Sachdev,
  ``{Thermoelectric transport in disordered metals without quasiparticles: The
  Sachdev-Ye-Kitaev models and holography},''
  \href{http://dx.doi.org/10.1103/PhysRevB.95.155131}{{\em Phys. Rev.}
  {\bfseries B95} no.~15, (2017) 155131},
\href{http://arxiv.org/abs/1612.00849}{{\ttfamily arXiv:1612.00849
  [cond-mat.str-el]}}.

\bibitem{Sachdev:1992fk}
S.~Sachdev and J.~Ye, ``{Gapless spin fluid ground state in a random, quantum
  Heisenberg magnet},''
  \href{http://dx.doi.org/10.1103/PhysRevLett.70.3339}{{\em Phys. Rev. Lett.}
  {\bfseries 70} (1993) 3339},
\href{http://arxiv.org/abs/cond-mat/9212030}{{\ttfamily arXiv:cond-mat/9212030
  [cond-mat]}}.

\bibitem{parcollet}
O.~Parcollet and A.~Georges, ``{Non-Fermi-liquid regime of a doped Mott
  insulator},'' \href{http://dx.doi.org/10.1007/JHEP04(2010)019}{{\em
  Phys.Rev.} {\bfseries B59} (1999) 019},
\href{http://arxiv.org/abs/9806119}{{\ttfamily arXiv:9806119 [cond-mat]}}.

\bibitem{Coleman:1969sm}
S.~R. Coleman, J.~Wess, and B.~Zumino, ``{Structure of phenomenological
  Lagrangians. 1.},''
\href{http://dx.doi.org/10.1103/PhysRev.177.2239}{{\em Phys. Rev.} {\bfseries
  177} (1969) 2239--2247}.

\bibitem{Callan:1969sn}
C.~G. Callan, Jr., S.~R. Coleman, J.~Wess, and B.~Zumino, ``{Structure of
  phenomenological Lagrangians. 2.},''
\href{http://dx.doi.org/10.1103/PhysRev.177.2247}{{\em Phys. Rev.} {\bfseries
  177} (1969) 2247--2250}.

\bibitem{Volkov:1973vd}
D.~V. Volkov, ``{Phenomenological Lagrangians},''
{\em Fiz. Elem. Chast. Atom. Yadra} {\bfseries 4} (1973) 3--41.

\bibitem{Ivanov:1975zq}
E.~A. Ivanov and V.~I. Ogievetsky, ``{The Inverse Higgs Phenomenon in Nonlinear
  Realizations},''
\href{http://dx.doi.org/10.1007/BF01028947}{{\em Teor. Mat. Fiz.} {\bfseries
  25} (1975) 164--177}.

\bibitem{Low:2001bw}
I.~Low and A.~V. Manohar, ``{Spontaneously broken space-time symmetries and
  Goldstone's theorem},''
  \href{http://dx.doi.org/10.1103/PhysRevLett.88.101602}{{\em Phys. Rev. Lett.}
  {\bfseries 88} (2002) 101602},
\href{http://arxiv.org/abs/hep-th/0110285}{{\ttfamily arXiv:hep-th/0110285
  [hep-th]}}.

\bibitem{Anninos:2015eji}
D.~Anninos, F.~Denef, and R.~Monten, ``{Grassmann Matrix Quantum Mechanics},''
  \href{http://dx.doi.org/10.1007/JHEP04(2016)138}{{\em JHEP} {\bfseries 04}
  (2016) 138},
\href{http://arxiv.org/abs/1512.03803}{{\ttfamily arXiv:1512.03803 [hep-th]}}.

\bibitem{Banerjee:2016ncu}
S.~Banerjee and E.~Altman, ``{Solvable model for a dynamical quantum phase
  transition from fast to slow scrambling},''
  \href{http://dx.doi.org/10.1103/PhysRevB.95.134302}{{\em Phys. Rev.}
  {\bfseries B95} no.~13, (2017) 134302},
\href{http://arxiv.org/abs/1610.04619}{{\ttfamily arXiv:1610.04619
  [cond-mat.str-el]}}.

\bibitem{Anninos:2017hhn}
D.~Anninos and D.~M. Hofman, ``{Infrared Realization of dS$_2$ in AdS$_2$},''
\href{http://arxiv.org/abs/1703.04622}{{\ttfamily arXiv:1703.04622 [hep-th]}}.

\bibitem{Cotler:2016fpe}
J.~S. Cotler, G.~Gur-Ari, M.~Hanada, J.~Polchinski, P.~Saad, S.~H. Shenker,
  D.~Stanford, A.~Streicher, and M.~Tezuka, ``{Black Holes and Random
  Matrices},'' \href{http://dx.doi.org/10.1007/JHEP05(2017)118}{{\em JHEP}
  {\bfseries 05} (2017) 118},
\href{http://arxiv.org/abs/1611.04650}{{\ttfamily arXiv:1611.04650 [hep-th]}}.

\bibitem{Chen:2017yze}
H.~Chen, C.~Hussong, J.~Kaplan, and D.~Li, ``{A Numerical Approach to Virasoro
  Blocks and the Information Paradox},''
\href{http://arxiv.org/abs/1703.09727}{{\ttfamily arXiv:1703.09727 [hep-th]}}.

\bibitem{Anninos:2016klf}
D.~Anninos and G.~A. Silva, ``{Solvable Quantum Grassmann Matrices},''
  \href{http://dx.doi.org/10.1088/1742-5468/aa668f}{{\em J. Stat. Mech.}
  {\bfseries 1704} no.~4, (2017) 043102},
\href{http://arxiv.org/abs/1612.03795}{{\ttfamily arXiv:1612.03795 [hep-th]}}.

\bibitem{Kunduri:2013ana}
H.~K. Kunduri and J.~Lucietti, ``{Classification of near-horizon geometries of
  extremal black holes},'' \href{http://dx.doi.org/10.12942/lrr-2013-8}{{\em
  Living Rev. Rel.} {\bfseries 16} (2013) 8},
\href{http://arxiv.org/abs/1306.2517}{{\ttfamily arXiv:1306.2517 [hep-th]}}.

\bibitem{Maldacena:1997ih}
J.~M. Maldacena and A.~Strominger, ``{Universal low-energy dynamics for
  rotating black holes},''
  \href{http://dx.doi.org/10.1103/PhysRevD.56.4975}{{\em Phys. Rev.} {\bfseries
  D56} (1997) 4975--4983},
\href{http://arxiv.org/abs/hep-th/9702015}{{\ttfamily arXiv:hep-th/9702015
  [hep-th]}}.

\bibitem{Denef:2007vg}
F.~Denef and G.~W. Moore, ``{Split states, entropy enigmas, holes and halos},''
  \href{http://dx.doi.org/10.1007/JHEP11(2011)129}{{\em JHEP} {\bfseries 11}
  (2011) 129},
\href{http://arxiv.org/abs/hep-th/0702146}{{\ttfamily arXiv:hep-th/0702146
  [hep-th]}}.

\bibitem{Bredberg:2009pv}
I.~Bredberg, T.~Hartman, W.~Song, and A.~Strominger, ``{Black Hole
  Superradiance From Kerr/CFT},''
  \href{http://dx.doi.org/10.1007/JHEP04(2010)019}{{\em JHEP} {\bfseries 04}
  (2010) 019},
\href{http://arxiv.org/abs/0907.3477}{{\ttfamily arXiv:0907.3477 [hep-th]}}.

\bibitem{Anninos:2009jt}
D.~Anninos, ``{Sailing from Warped AdS(3) to Warped dS(3) in Topologically
  Massive Gravity},'' \href{http://dx.doi.org/10.1007/JHEP02(2010)046}{{\em
  JHEP} {\bfseries 02} (2010) 046},
\href{http://arxiv.org/abs/0906.1819}{{\ttfamily arXiv:0906.1819 [hep-th]}}.

\bibitem{Pioline:2005pf}
B.~Pioline and J.~Troost, ``{Schwinger pair production in AdS(2)},''
  \href{http://dx.doi.org/10.1088/1126-6708/2005/03/043}{{\em JHEP} {\bfseries
  03} (2005) 043},
\href{http://arxiv.org/abs/hep-th/0501169}{{\ttfamily arXiv:hep-th/0501169
  [hep-th]}}.

\bibitem{Booth:1998gf}
I.~S. Booth and R.~B. Mann, ``{Cosmological pair production of charged and
  rotating black holes},''
  \href{http://dx.doi.org/10.1016/S0550-3213(98)00756-1}{{\em Nucl. Phys.}
  {\bfseries B539} (1999) 267--306},
\href{http://arxiv.org/abs/gr-qc/9806056}{{\ttfamily arXiv:gr-qc/9806056
  [gr-qc]}}.

\bibitem{Anninos:2009yc}
D.~Anninos and T.~Hartman, ``{Holography at an Extremal De Sitter Horizon},''
  \href{http://dx.doi.org/10.1007/JHEP03(2010)096}{{\em JHEP} {\bfseries 03}
  (2010) 096},
\href{http://arxiv.org/abs/0910.4587}{{\ttfamily arXiv:0910.4587 [hep-th]}}.

\bibitem{Anninos:2010gh}
D.~Anninos and T.~Anous, ``{A de Sitter Hoedown},''
  \href{http://dx.doi.org/10.1007/JHEP08(2010)131}{{\em JHEP} {\bfseries 08}
  (2010) 131},
\href{http://arxiv.org/abs/1002.1717}{{\ttfamily arXiv:1002.1717 [hep-th]}}.

\end{thebibliography}\endgroup

\end{document}